\documentclass[10pt, twocolumn, floatfix, superscriptaddress, aps]{revtex4-2}
\usepackage{amsmath, amssymb, graphicx, xcolor, colortbl, braket}
\usepackage{graphicx}
\usepackage{epsfig}
\usepackage{makecell}
\usepackage{dcolumn}
\usepackage{bm}
\usepackage{amssymb}
\usepackage{amsmath}
\usepackage{ragged2e}
\usepackage{upgreek}
\usepackage{tabularx}
\usepackage{enumitem}  
\usepackage{float}
\usepackage{lineno}
\usepackage[colorlinks,citecolor=blue,linkcolor=blue,hyperindex]{hyperref}
\usepackage{hyperref}
\hypersetup{
  linkcolor=blue,
  urlcolor=blue
}

\newcommand{\addt}[1]{\textcolor{black}{#1}}
\newcommand{\addtt}[1]{\textcolor{black}{#1}}

\begin{document}

\title{Haerter-Shastry kinetic magnetism and metallicity in the triangular Hubbard model} 
\author{Sogoud Sherif}
\email{ssherif@fsu.edu}
\affiliation{National High Magnetic Field Laboratory, Tallahassee, Florida 32310, USA}
\affiliation{Department of Physics, Florida State University, Tallahassee, Florida 32306, USA}
\author{Prakash Sharma}
\affiliation{Department of Chemistry, Emory University, Atlanta, Georgia 30322, USA}
\author{Aman Kumar}
\affiliation{National High Magnetic Field Laboratory, Tallahassee, Florida 32310, USA}
\affiliation{Department of Physics, Florida State University, Tallahassee, Florida 32306, USA}
\author{Hitesh J. Changlani}
\email{Corresponding author: hchanglani@fsu.edu}
\affiliation{National High Magnetic Field Laboratory, Tallahassee, Florida 32310, USA}
\affiliation{Department of Physics, Florida State University, Tallahassee, Florida 32306, USA}


\begin{abstract}
\noindent

\addtt{The fermionic Hubbard model, when combined with frustration, associated with the breaking of particle-hole symmetry, harbors a rich phase diagram. Theoretical findings associated with the nature of magnetism and metallicity across this phase diagram are now being actively investigated in triangular Hubbard cold atom and solid-state (moiré) based emulators. Building on the theoretical work of Haerter and Shastry, we explore the impact of kinetically frustrated magnetism, a phenomenon where antiferromagnetic order emerges without any underlying magnetic interactions, at finite hole density. We numerically study the infinite-$U$ triangular Hubbard model using the density matrix renormalization group algorithm and estimate the stability of the kinetically induced $120^{\circ}$ antiferromagnetic state to hole doping. Beyond the Haerter-Shastry regime, we find an intermediate phase with multimer (involving multiple correlated spins) stripes that eventually gives way to a paramagnet. We also find evidence of gapless charge excitations (metallicity) throughout the phase diagram for finite hole density. We discuss the implications at large, but finite, values of $U/t$, and investigate whether kinetic magnetism and superexchange collaborate or compete.}
\end{abstract}
{\maketitle}

\noindent
\MakeUppercase{\textbf{Introduction}}\\
Can magnetism emerge in the absence of magnetic interactions? The answer to this somewhat paradoxical question is now known to be ``yes"-- with the mechanism rooted in the interplay of strong interactions and spin degrees of freedom. Exemplifying this is the Nagaoka (or Nagaoka-Thouless) ferromagnet \cite{NagaokaFM, Thouless_1965}-- on the square and other bipartite lattices (and even non-bipartite lattices, for the hopping $t<0$.) in the limit of extreme interactions, a single hole in a half-filled lattice 
favors a ferromagnetic (FM) background as a way of minimizing its energy. 

The framework for addressing these questions is the $t-J$ model~\cite{Spalek_1978}, 
\begin{flalign}
    &H_{t-J} = -t\sum_{\langle i,j \rangle,\sigma}(\mathcal{P} c_{i,\sigma}^\dagger c_{j,\sigma}\mathcal{P} + \textrm{h.c.})+ J\sum_{\langle i,j \rangle}(\vec{S_{i}}\cdot \vec{S_{j}}-\frac{n_{i}n_{j}}{4})&
    \label{eq:tJmodel}
\end{flalign} 
which is the low-energy limit of the Hubbard model~\cite{Hubbard_1963},
\addt{
\begin{flalign}
    &H_{Hubbard} = -t\sum_{\langle i,j \rangle,\sigma}(c_{i,\sigma}^\dagger c_{j,\sigma} + \textrm{h.c.})+ U \sum_{i} n_{i,\uparrow}n_{i,\downarrow}&
    \label{eq:Hubbard}
\end{flalign}
}
where $c_{i,\sigma}^\dagger$ and $c_{i,\sigma}$ refer to the usual fermionic creation and annihilation operators respectively, with spin $\sigma=\{\uparrow,\downarrow\}$ at site
$i$, $n_{i,\sigma}=c_{i,\sigma}^\dagger c_{i,\sigma}$ and $n_i = n_{i,\uparrow}+n_{i,\downarrow}$ are number operators. $\langle i,j\rangle$ refer to nearest neighbor pairs, and
$\mathcal{P} = \prod_i(1-n_{i,\uparrow}n_{i,\downarrow})$ is a projector eliminating states involving double occupancy at any site.
 \addt{$t$ is the hopping, $U$ is the onsite Coulomb repulsion} and $J = 4t^2/U$ \addt{is} the effective Heisenberg exchange. (Throughout this paper we will be concerned with $t>0$ and often set it to $t=1$.) \addtt{For} $U/t = \infty$, $J$ is exactly zero, hence no magnetic order is intrinsically favorable to begin with; and any selection of magnetic order stems purely from the kinetic term \addt{in Eq.~\eqref{eq:tJmodel}}. While the Nagaoka theorem (and its extension~\cite{Tasaki_1989}) serves as a guiding principle for such state selection, it typically does not apply in situations with geometric frustration, or for more than one hole. In such cases, the nature of the selected magnetic state is open-ended, and thus has been the subject of considerable investigation (see, for example, ~\cite{Roth_1969,Doucot_PRB_1989,ska, Basile_Elser, Hanisch_PRB_1997, Becca_PRL_2001, Haerter_PRL_2005, Yin2011, sposetti, lisandrini2017evolution,Pereira_arXiv_2025, Sharma_arxiv_2025,zhang2018pairing,glittum2025resonant,ferro_dmrg_triangular,prichard2024directly,lebrat2024observation,koepsell2019imaging} and references therein).

\begin{figure}
     \includegraphics[width=0.5\textwidth]{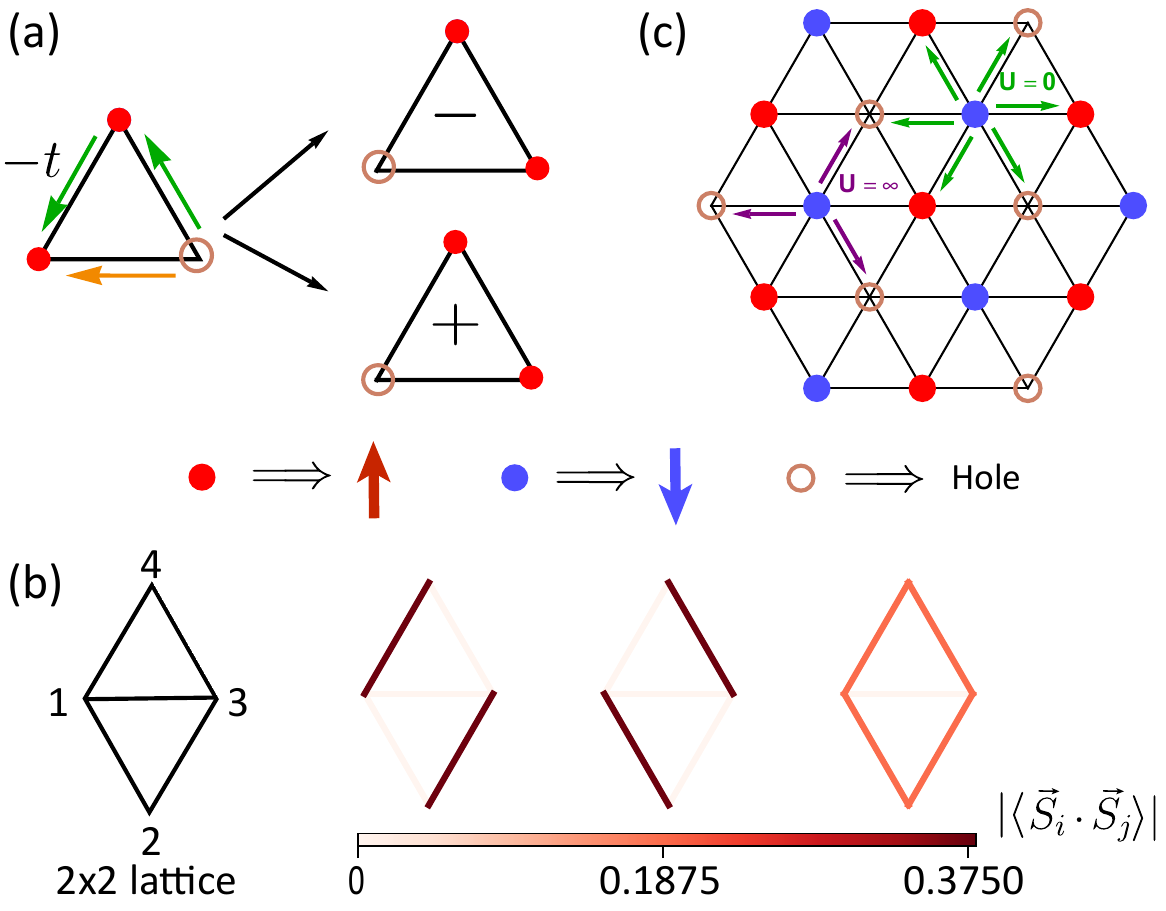}
    \caption{\textbf{\addtt{Hole dynamics on the triangular Hubbard model}} (a) Schematic of a single hole hopping on a triangular plaquette in the classical \addtt{ferromagnetic} background via two available routes, represented by green and orange arrows. The hole's motion along the two paths interferes destructively. (b) The $2\times 2$ triangular unit cell with periodic boundary conditions, \addt{where the connection between sites 2 and 4 is not shown} (left) and the absolute value of ground state spin-spin correlations on the bonds for three eigensolutions for one hole, two up spins and one down spin (right). (c) Schematic showing available hopping pathways for the $U=0$ and $U=\infty$ models.}
    \label{fig:fig1}
\end{figure}

\begin{figure*}
    \centering
        \includegraphics[clip, trim=3.4cm 0cm 3.4cm 1.3cm, width=\textwidth]{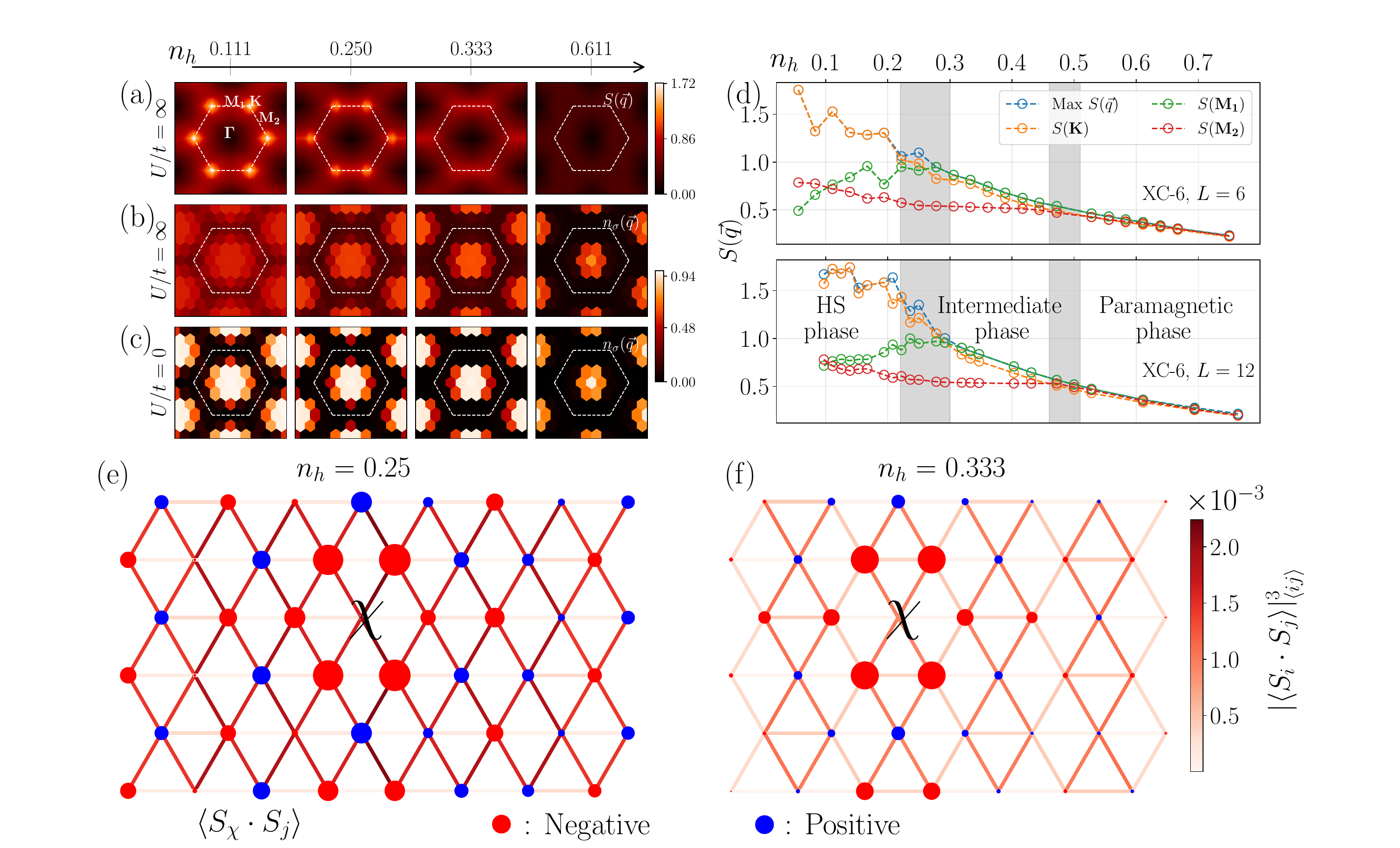}
    \caption{\justifying \textbf{\addtt{Magnetic and metallic properties induced by hole doping the triangular Hubbard model}} (a) Ground state static spin structure factor $S(\vec{q})$, \addtt{as in Eq.\eqref{eq:spinstrfact}}, of the 72-site (XC-$6$ cylinder of length \addtt{$L=12$}, restricted to sites in the bulk), triangular Hubbard model for $U/t = \infty$ across a representative sample of hole \addt{densities} $n_{h}$.
    (b-c) Ground state momentum space fermionic occupation number of one spin species $n_{\sigma}(\vec{q})$, \addtt{as in Eq.\eqref{eq:mom_occ}}, for (b) $U/t = \infty$ and (c) $U/t = 0$. 
    For (a)-(c) the horizontal and vertical axes
correspond to $q_x$ and $q_y$ respectively, and the center corresponds to $(0,0)$. The hexagon, indicated by the white dashed line, marks
the boundary of the first Brillouin zone of the triangular lattice.  
    (d) Bulk static spin structure factor at
    representative momentum space points [$\mathbf{K}$ , $\mathbf{M}_{1}$ and $\mathbf{M}_{2}$ shown in (a)] as a function of  $n_h$. (e-f) Real space profile of spin correlations $\langle \vec{S_i}\cdot \vec{S_j}\rangle$ \addtt{for} (e) $n_h=1/4$ and (f) $n_h=1/3$, \addtt{on the XC-$6$ cylinder of length $L=9$, restricted to sites in the bulk}. The red color variation on the bonds represents nearest-neighbor AFM correlations, while the color (red, negative and blue, positive) and size of the dots indicate the expectation value of the spin-spin correlations with respect to a reference site $\chi$.}
    \label{fig:fig2}
\end{figure*}
In this paper, we study the hole-doped infinite-$U$ triangular Hubbard model (ITHM for short) and map out its phase diagram using the density matrix renormalization group (DMRG) algorithm~\cite{White_1992, Stoudenmire_2012}. Our analysis builds on pioneering work by Haerter and Shastry (HS), who showed that a single hole in the ITHM favors a $120^{\circ}$ \addtt{antiferromagnetic} (AFM) \addtt{spin} background~\cite{Haerter_PRL_2005}, in sharp contrast to the FM in the unfrustrated case. Note that a single hole alone can not destroy magnetic order associated with macroscopically many degrees of freedom in realistic conditions, where $U/t$ is large but finite. With this motivation, we explore the case of finite hole densities and examine what effects persist for large, but finite, $U/t$ on the hole-doped side
(see work on the finite $U/t$ phase diagram in Refs.~\cite{Li2014entropy, merino2006ferromagnetism,yoshioka2009quantum,sahebsara2008hubbard,wietek2021mott,beyond_hesinberg2017,prichard2024directly} and references therein). 

More generally, we expect the ITHM to be a valid effective model in the intermediate temperature regime, above the magnetic ordering scale but well below the Hubbard $U$ scale, where double occupancy physics is completely suppressed, and kinetic effects dominate i.e. $t^2/U \lesssim T \lesssim U$~\cite{lee2023triangular, Morera_2023}. But at $T=0$ it is \textit{a priori} not clear what aspects of the ITHM persist to finite $U/t$, especially at low density of holes. Our work clarifies the regimes in which the physics of superexchange crosses over to HS kinetic frustration, and we make concrete predictions for its observation in future experiments. 
Our findings are relevant to recently realized triangular Hubbard emulators (cold atoms~\cite{greinerXu,mazurenko2017cold,koepsell2019imaging,qiao2025kineticallyinducedboundstatesfrustrated}, optical lattices \cite{lebrat2024observation}, and solid state moiré systems~\cite{tang, ciorciaro2023kinetic, tao2024observation, Carlstrom_2022} \addt{created from transition metal dichalcogenide layers}) that have been studied for a wide range of hole densities at finite temperature. 
\\
\\
\noindent
\MakeUppercase{\textbf{Results and Discussion}}
\\
{\bf{Origin of antiferromagnetism on doping in the presence of frustration}}\\
The origin of antiferromagnetic tendencies on doping can already be seen on small local motifs of few sites; we provide exact solutions in \addtt{Methods -- Exact solution for one hole on small motifs}, to support our assertions below. 
Consider a single triangle with a single hole and two fully polarized spins. The hole can hop to a neighboring 
site via two different paths shown in Fig.~\ref{fig:fig1}\textcolor{blue}{(a)} (phases from fermionic signs are shown, also see ref. \cite{Sharma_arxiv_2025}); the amplitudes associated with 
these two paths interfere destructively, effectively reducing the energetic gain associated with the delocalization of the hole; this has 
been previously identified as the essence of kinetic frustration~\cite{barford1991spinless}. The ground state is two-fold degenerate, with energy $-t$. In contrast, on a square plaquette with three fully polarized spins and one hole, the analogous interference  
is constructive and leads to a unique ground state with energy $-2t$. 

When the two spins on the triangle are opposite, there is no such destructive interference, as they lead to two distinct configurations. 
In this case, the ground state is non-degenerate, its energy is $-2t$, i.e., lower than the FM state. This suggests that hole motion is enhanced by local antiferromagnetic correlations, highlighting the role of frustration in reshaping the magnetic responses to doping. For the case of 2 up spins and 1 down spin on the square plaquette,
the energy is identical to that of the FM, SU(2) symmetry dictates its wavefunction is $S^{-}|FM\rangle$; and the result is consistent with the Nagaoka condition. \addtt{(The square plaquette is topologically equivalent to a 4-site one
  dimensional chain. Strictly speaking, Nagaoka's ergodicity condition, required for the
  Perron Frobenius theorem, breaks down in one-dimension and thus it is not guaranteed that the
  ground state is unique. We ignore this subtlety for the illustrative argument
  here.)}

Now consider frustrating the square plaquette with diagonal bonds -- this corresponds to a $2 \times 2$ triangular unit cell with periodic boundary conditions [Fig.~\ref{fig:fig1}\textcolor{blue}{(b)}]. For one hole and 2 up spins and 1 down spin (or 2 down spins and 1 up spin), there are three exactly degenerate ground states with energy $-2t$. These correspond to states that resemble the three ``dimer coverings" of the $2\times 2$ lattice -- we have shown spin-spin correlations for two of these, and have not shown the covering where sites 1,3 and 2,4 pair up. The exact solution reveals that each one of these is not a product state -- instead, it is a superposition of states where a singlet is formed on a pair of sites, while the remaining up spin is delocalized on the other two sites. An equally weighted linear combination of the two states shown, results in spin correlations that appear as a ``diamond". We show later that
competing tendencies from such degenerate states qualitatively carry over to the case of hole density $n_h = 1/4$ on bigger system sizes.

To understand the true nature of the antiferromagnetic state in the thermodynamic limit, especially at small $n_h$, requires looking beyond local motifs.
The case of a single hole on an extended triangular lattice was addressed by HS, who derived an effective (non-local) spin Hamiltonian and then used exact diagonalization and a correspondence to the (local) triangular Heisenberg antiferromagnet~\cite{Haerter_PRL_2005}. The results suggested the existence of a ``tower of states" in the many-body spectrum~\cite{Bernu_1992}, associated with symmetry breaking, thereby providing evidence for the occurrence of the $120^{\circ}$ AFM ground state. This was subsequently verified and developed further with the help of DMRG and interpreted within the framework of slave fermion mean-field theory~\cite{sposetti}.
From the lens of a single hole effectively hopping in the background of spins, the latter provide the necessary Berry phases for the hole to lower its energy. 
\\
\\
\noindent
{\bf{Magnetism due to kinetic frustration at finite hole density}}\\
To extend our analyses beyond a single hole, we have carried out large-scale DMRG
calculations for ITHM across various cylindrical geometries (\addtt{see Methods --  Analysis of geometry and finite-size dependence of numerical results}) and for a wide range of hole densities $n_h \equiv N_h/N$ on top of half-filling, for $N$ sites and 
$N_{h}$ holes. Our main conclusions are based on width 6 cylinders with the XC boundary condition (in which one \addt{edge of the triangles} runs parallel to the length \addtt{$(L)$} of the cylinder and the boundaries are zigzag bonds,  
and we use bond dimensions for the underlying matrix product state to 
be as large as $m=14000$, which ensures a truncation error below $1.6 \times 
10^{-5}$. We restrict our calculation to the $S_z=0$ \addt{sector} for an even number of holes 
and $S_z=1/2$ for an odd number of holes. Typically, cylinder sizes and boundary conditions that accommodate $120^{\circ}$ AFM order were chosen. 

We compute the spin structure factor in the ground state,
\begin{equation}
S(\vec{q}) \equiv \frac{1}{N}\sum_{i,j} e^{i \vec{q} \cdot (\vec{r}_i - \vec{r}_j)} \langle \vec{S_i} \cdot \vec{S_j} \rangle.
\label{eq:spinstrfact}
\end{equation} 
To minimize boundary effects, we have used two different strategies. In the first strategy, we cut out sites from the boundary, and assuming their number is $N_b$, the $N$ in the expression is replaced with $N - N_b$ and the summation is carried out only for the retained (``bulk") sites. We refer to this as ``method 1". In the second strategy (``method 2"), we consider a reference site ($\chi$) close to the center of the finite sample, and compute $S(\vec{q}) \equiv \sum_{j} e^{i \vec{q} \cdot (\vec{r}_\chi -\vec{r}_j)} \langle \vec{S_{\chi}} \cdot \vec{S_j} \rangle$. For a completely translationally invariant ground state, both strategies are exactly equivalent. We find them to be in reasonable agreement despite
the limitation posed by the cylindrical geometry, with the notable exception of
the case of very low hole density, as we explain shortly and also revisit in \addtt{Methods -- Analysis of antiferromagnetic ordering in the Haerter- Shastry regime}. For the case of one hole, we observe that $S(\vec{q})$ peaks at $(\frac{4\pi}{3},0)$ and other symmetry-related $\mathbf{K}$ points, consistent with the $120^{\circ}$ AFM order, confirming the findings of previous work~\cite{Haerter_PRL_2005,sposetti}. 

We also compute $n_{\sigma}(\vec{q})$; the momentum space occupation of one spin species of electron,
\begin{equation}
n_{\sigma}(\vec{q}) \equiv \frac{1}{N} \sum_{i,j} e^{i \vec{q} \cdot (\vec{r}_i - \vec{r}_j)} \langle c^{\dagger}_{i,\sigma} c_{j,\sigma}\rangle. 
\label{eq:mom_occ}
\end{equation}
Results of both quantities for representative hole densities are shown in Fig.~\ref{fig:fig2}\textcolor{blue}{(a,b)}. For comparison, we also show the non-interacting ($U=0$) $n_{\sigma}(\vec{q})$, for the same density and cylinder in Fig.~\ref{fig:fig2}\textcolor{blue}{(c)}. 

In the low-density limit of holes, the HS picture of the $120^{\circ}$ AFM qualitatively holds; $S(\vec{q})$ peaks at the $\mathbf{K}$ points, and the order parameter is non-zero (see 
\addtt{Methods -- Analysis of antiferromagnetic ordering in the
Haerter-Shastry regime}), consistent with long-range AFM. As $n_h$ increases, the weight at the $\mathbf{K}$ points diminishes 
and spreads out, indicating the onset of an intermediate phase, which breaks the six-fold rotational symmetry. (The choice of cluster also breaks this symmetry, and so the precise location and extent of this phase is sensitive to finite size effects). The six-fold symmetry is then (approximately) restored at higher $n_h$, where the role of the interaction strength is less important, and a paramagnetic phase is observed, characterized by a featureless $S(\vec{q})$ [Fig.~\ref{fig:fig2}\textcolor{blue}{(a)}]. To get the approximate locations of these phases, we monitor the weight at other points and look for where key features begin to change, e.g. $S(\bf K) \approx S(\bf M_1)$ or $S{(\bf M_1)} \approx S{(\bf M_2)}$. While this is rigorous (in the thermodynamic limit) for the HS phase where $S(\bf K)$ is the order parameter, it only serves as a guide for whether the putative intermediate phase exists. \addt{[While we can not rule out that the intermediate and paramagnetic phases are related by a crossover, we find their short range real spin correlations to be strikingly different. We discuss this observation shortly].}

In Fig.~\ref{fig:fig2}\textcolor{blue}{(d)} we plot $S(\vec{q})$, using method 1, as a function of $n_h$ for representative momenta, for two system sizes, to determine the approximate location of the phase boundaries.
The extent of the HS $120^{\circ}$ AFM up to $n_{h}\simeq0.2-0.3$ is similar to that reported for the Nagaoka FM for the square case~\cite{squareinfU}, where a commensurate ``plaquette insulator" at $n_h=1/4$ has been reported to lead to a region of phase separation for $0.2 \leq n_h < 0.25$. The local motifs (square plaquettes) in this insulator each individually harbor a Nagaoka FM state, each acting like an effective spin 3/2 moment, that are weakly coupled to one another via effective AFM interactions. 
This naturally raises the question whether an analog exists for the ITHM, and whether it is at the core of the intermediate phase we see.

The spin structure factor is useful to infer global properties of the magnetic state favored, but is relatively insensitive to local physics, which can be in the form of singlet/valence bond formation. Thus, to gain further insight, we view our findings in real space in Fig.~\ref{fig:fig2}\textcolor{blue}{(e)} and ~\ref{fig:fig2}\textcolor{blue}{(f)} where we focus our attention on specific densities, $n_h=1/4$ and $n_h=1/3$ respectively. We plot the spin-spin correlations in the ground state $\langle \vec{S}_i \cdot \vec{S}_j\rangle$ with respect to a reference site $\chi$, the magnitude and sign on a site is indicated by the size and color of the circle on that site (red is negative, blue is positive). We find a rapid decay in the $x-$ (long) direction and sizable value in the $y-$ (short) direction. The Fourier transform of such a pattern of correlations is essentially featureless along $q_x$, yet confined to a small window of $q_y$, explaining our observations in Fig. \ref{fig:fig2}\textcolor{blue}{(a)} in the intermediate $n_h$ regime. 

We also examine the real-space spin-spin correlations between neighboring sites, whose strengths are indicated by the color intensity of the red bonds, all of which are found to be antiferromagnetic, but with important modulations. For presentational purposes and to highlight the differences between stronger and weaker correlations, we have plotted the cube of this value.
We also eliminate the boundary spins and bonds to highlight the bulk physics. For small $n_h$ the DMRG calculations are difficult to converge to high precision, despite the large bond dimension employed, leading to the appearance of trapped holes or modulated nearest neighbor correlations. \addt{While the physics associated with holes/charge degrees of freedom will be discussed extensively in the next section, we clarify here that such behavior} is not entirely unexpected -- the motion of the holes is frustrated, they are heavy, and many competitive spin states exist in the low-energy landscape. However, on average, the real space spin-spin correlations are long-ranged and consistent with $120^\circ$ order, which results in the structure factor pattern shown in Fig.~\ref{fig:fig2}\textcolor{blue}{(a)}. A finite-size analysis of the order parameter based on using $S(\mathbf{K})$ further confirms the long-range nature of the magnetic order (\addtt{see Methods -- Analysis of antiferromagnetic ordering in the Haerter-Shastry regime}).

Remarkably, for the case of $n_h = 1/4$, the ``diamond type" spin correlations that we previously noted for a $2 \times 2$ cluster \addtt{(see Methods -- Exact solution for one hole on small motifs)}, are robustly seen throughout the entire cluster, as is shown in Fig.~\ref{fig:fig2}\textcolor{blue}{(e)}. While there is some migration of particles towards the left and right boundaries of the cylinder, the hole density in the bulk is close to $1/4$. 
\addt{The exact solution for one hole on the $2 \times 2$ triangular cluster reveals an exact three-fold ground state degeneracy, of which the diamond pattern corresponds to one possible linear combination, thus it is conceivable that other patterns may be stabilized on other finite systems. We find that this is indeed the case -- } 
the nature of spin multimer stabilized is sensitive to the system size and boundary conditions -- in \addtt{Methods -- Analysis of geometry and finite-size dependence of numerical results,} we show some examples where other states are seen in the DMRG calculations. In the case of $n_h=1/3$, as in Fig.~\ref{fig:fig2}\textcolor{blue}{(f)}, we observe an overall weakening of these nearest-neighbor antiferromagnetic correlations throughout the cylinder, yet prominent \addtt{additional} modulations emerge. Specifically, the diamond-type local spin correlations for $n_h=1/4$ persist along the vertical direction in columns forming ``stripes", but the stripes themselves are now weakly correlated to one another in the $x$-direction. The correlations with respect to a reference spin show similar short-range and anisotropic character as the case of $n_h=1/4$, and thus from the point of view of the structure factor, many features appear continuously related. Finally, for large hole densities, all correlations are significantly weakened, consistent with the onset of a paramagnetic phase. 

Thus the overall emergent picture from our calculations is that the HS 120$^{\circ}$ \addtt{AFM} melts to a paramagnet via an intermediate phase, which is characterized by local multi-spin correlations.
\addtt{(A short-range AFM flanking the long-range ordered 120$^{\circ}$ AFM has been previously suggested in the framework of dynamical mean field theory (DMFT)~\cite{Li2014entropy} and DMRG calculations~\cite{lee2023triangular} at finite, but large $U/t$ -- however its true nature remains unclear. In the current setup, superexchange is completely absent.)} 
This intermediate phase melts by first decoupling into stripes, which eventually melt internally as well, leading to the paramagnet.\\

\begin{figure}
\includegraphics[clip, trim=0.3cm 0.25cm 0.25cm 0.2cm,width=\linewidth]{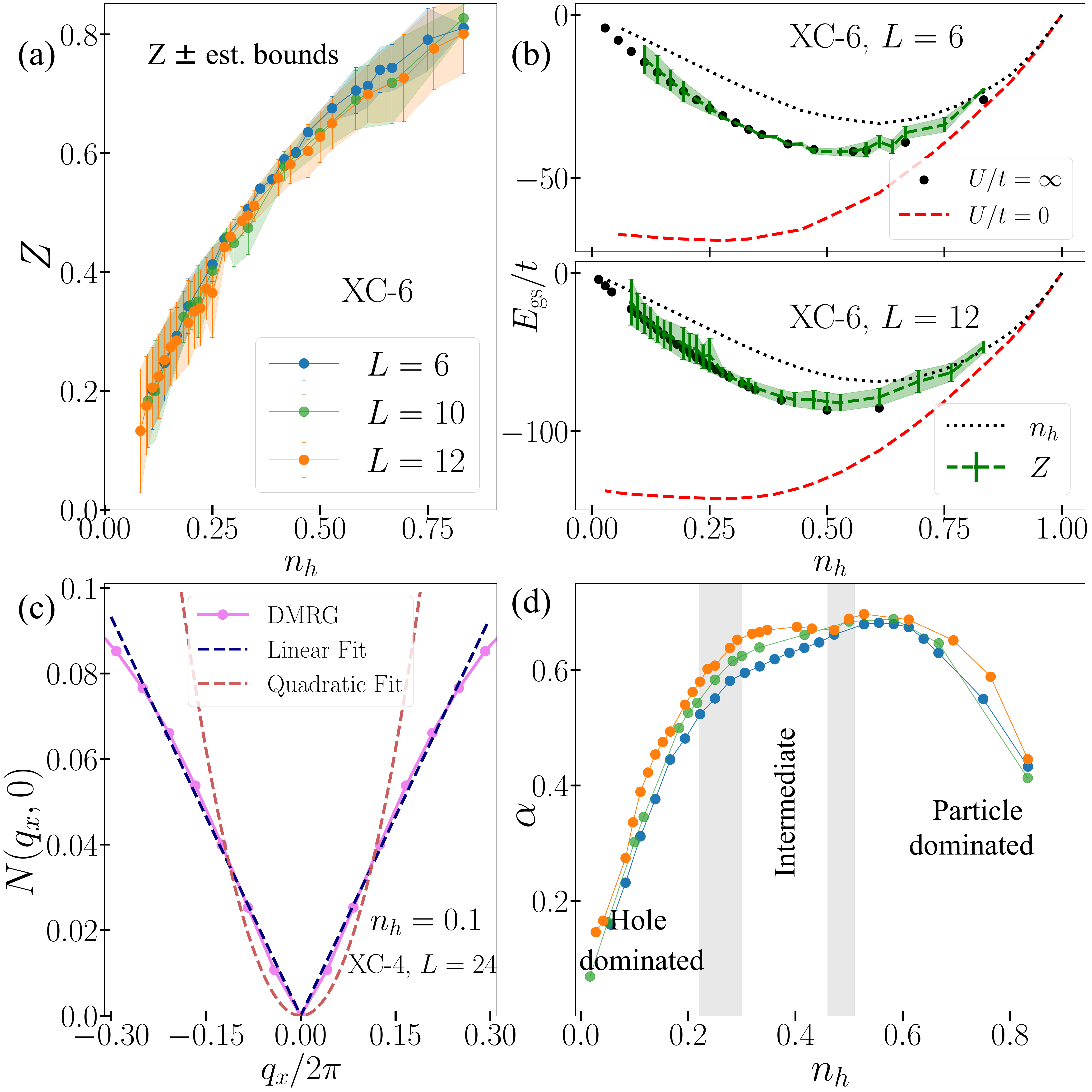}
\caption{\justifying \textbf{\addtt{Characterization of the metallic properties of the doped triangular Hubbard model}} (a) Quasiparticle weight \addt{$(Z)$}, \addtt{as in Eq.\eqref{eq: qw}},
versus hole density ($n_h$), computed from the \addtt{$U/t=\infty$} ground state wavefunction for three representative system sizes \addt{all of type XC-6 \addtt{with length $L$}}. \addtt{The difference in the largest and smallest occupations in momentum space $n_{\sigma}(\vec{q})$, \addtt{as in Eq.\eqref{eq:mom_occ}}, provides estimated upper and lower bounds (est. bounds) on $Z$, which are used to obtain the uncertainty. (see Methods -- Determining quasiparticle weight).} (b) Ground state energies for $U/t=\infty$ (denoted by black dots), and $U/t=0$ (red dashed line) versus $n_h$, shown separately for \addt{the XC-6 \addtt{$L=6$ and $L=12$ cylinders}}. The dotted and black dashed lines represent the $U/t=0$ ground state energy rescaled by $n_h$ and $Z$ \addtt{(incorporating the uncertainty in $Z$ for $n_h > 0.1$ )}, respectively. (c) Ground state static charge structure factor $N(\vec{q})$, \addtt{as in Eq.\eqref{eq:ch_str_fac}}, for $q_{y}=0$ on the 96-site (XC-$4$ \addtt{$L=24$}) \addtt{for $U/t=\infty$} for $n_{h}=0.1$. The linear and quadratic fits close to $q_x\rightarrow 0$ are shown. 
(d) Metallic weight, $\alpha$, derived numerically from the linear fits of the $N(\vec{q})$, and as defined in Eq. \eqref{eq:alpha} , versus $n_{h}$ for the same systems as in (a). 
}
\label{fig:fig3}
\end{figure}

\noindent
{\bf{Metallic properties and energy renormalization}}
\\
We now focus our attention to the properties of the charge carriers themselves. At low hole densities, the effective quasiparticles (if they exist) are highly renormalized, arising from the kinetic obstructions for the holes. Since superexchange is completely suppressed, the pathways for hole delocalization are fewer, making them heavier. This manifests itself prominently in Fig. \ref{fig:fig2}\textcolor{blue}{(b)}, the momentum space occupation for $U = \infty$ at \addt{small} $n_h$ does not show a sharp signature of a Fermi jump, which is typically associated with an insulator, a non-Fermi liquid (nFL) or a highly dressed FL, in comparison to Fig. \ref{fig:fig2}\textcolor{blue}{(c)} of the occupation number at $U = 0$. At higher $n_h$, the visual similarity between the $U= \infty$ and $U=0$ limits is expectedly restored, and the origin of the anisotropy of the Fermi surface is predominantly from the choice of the shape of the cylinder chosen.

The extent of renormalization is precisely quantified with the quasiparticle weight (residue) $Z$~\cite{Coleman2015}, 
\begin{equation}
Z=n_{\sigma}(\vec{q}_{F}-(2\pi/L,0))-n_{\sigma}(\vec{q}_{F}+(2\pi/L,0))
  \label{eq: qw}
\end{equation}
where $\vec{q}_{F}$ is the Fermi momentum for the corresponding non-interacting tight-binding system.
[In Methods \addtt{(Determining quasiparticle weight)}, we discuss how $Z$ was extracted, with the help of some examples]. 
In Fig. \ref{fig:fig3}\textcolor{blue}{(a)}, we show how $Z$ varies with $n_h$, for three representative system sizes. While it is clear that $Z$ increases monotonically with increasing $n_h$; it is difficult to numerically establish whether or not it becomes zero at some small, but non-zero, $n_h$. 
Non-zero but small $Z$, at $T=0$, in the ITHM would mean a heavy, but conventional, FL ground state, which may crossover to a potential nFL at a small (on a scale of $t$) finite temperature. With reference to this observation, we remark that recent work has shown numerical evidence for $T$-linear resistivity, associated with nFL, on the infinite-$U$ square lattice Hubbard model at low hole density and low temperatures~\cite{Fratini_2025}. We expect the essence of this phenomenon to carry over to the ITHM, but a systematic investigation of the similarities and differences with the square case is left to future work. 

For larger $n_h$, where FL is naturally expected, quantitative differences still persist between the $U=\infty$ and $U=0$ cases [Figs.~\ref{fig:fig2}\textcolor{blue}{(b-c)}]. At low electron density, particles easily avoid each other, and the probability of double occupancy physics is suppressed in both cases. But as our cartoon picture in Fig.~\ref{fig:fig1}\textcolor{blue}{(c)} shows, the motion of the electrons is obstructed -- processes that would be allowed for $U=0$ are now completely forbidden for $U=\infty$. To a very crude approximation, the effective kinetic energy of the electrons is now directly proportionate to the density of holes -- the kinetic energy of the electrons is renormalized, i.e., the band is flatter, and this is reflected directly in the trend seen for $Z$. To quantify this further, we compare the ground state energy at $U=0$ and $U = \infty$ on the same finite cluster; the former is computed exactly, and the latter is obtained by an accurate DMRG calculation.
In Fig.~\ref{fig:fig3}\textcolor{blue}{(b)}, the $U=0$ ground state energy is rescaled by $n_h$, which \addt{gives the same trend as } 
the DMRG-computed energy at $U=\infty$, \addt{but is quantitatively inadequate}. Rescaling it by the numerically computed $Z$ achieves an \addt{excellent}
match, which supports the idea that there is metallic behavior (possibly conventional FL) throughout the entire phase diagram of the ITHM for $n_h > 0$. We find it remarkable that this simple rescaling works well in both hole-dominated and electron-dominated regimes.

Further evidence for gapless excitations for all phases of the ITHM is obtained by analyzing the charge structure factor $N(\vec{q})$, defined as
\begin{equation}
N(\vec{q}) = \frac{1}{N} \sum_{i,j} (\langle n_i n_j \rangle - \langle n_i \rangle \langle n_j \rangle ) e^{i\vec{q} \cdot (\vec{r}_i - \vec{r}_j) }
\label{eq:ch_str_fac}
\end{equation}
One can learn about insulating versus metallic behaviors from the scaling of $N(\vec{q})$ at small $|\vec{q}|$ \cite{feynman}. $N(\vec{q}) = \beta |\vec{q}|^2$ is characteristic of systems with a charge gap (i.e., insulators) \cite{Sorella} whereas for 
systems with gapless charge excitations (i.e., metals where the effective quasiparticle energy scales as $|\vec{q}|$) we expect,
\begin{equation}
    N(\vec{q}) = \alpha |\vec{q}|
    \label{eq:alpha}
\end{equation}
where the proportionality factor $\alpha$ is referred to as the metallic weight~\cite{wietek2021mott}. We clarify that in the context of DMRG, the (short) width direction does not provide a sufficient number of points to conclusively determine this scaling in that direction; instead, we show results only for the (long) length direction. A representative example for the \addtt{XC-4 $L=24$} cylinder with $n_h \approx 0.1$  \addt{(10 holes)} is shown in Fig.~\ref{fig:fig3}\textcolor{blue}{(c)}, showing an excellent fit to a straight line for small $|q_x|$, and considerable deviation from the quadratic form.  

We have observed the linear scaling of $N(\vec{q})$ with $q_x$ (fixing $q_y=0$) on the ITHM across all $n_h > 0$ that were studied, supporting the existence of gapless charge excitations. We plot $\alpha$ as a function of $n_h$ in Fig. \ref{fig:fig3}\textcolor{blue}{(d)}. 
[$Z$ should not be confused with $\alpha$ -- the former reveals the extent of the renormalization of the quasiparticles, the latter is an indirect measure of their density.] 
As shown, at small $n_h$, $\alpha$ increases with increasing $n_h$; which is consistent with holes acting as charge carriers in the HS regime.
On further doping, $\alpha$ appears to saturate - showing an almost flat feature in the region $ 0.25 \lesssim n_h \lesssim 0.6$, which coincides (roughly) with the location of the intermediate phase. Beyond this $n_h$, $\alpha$ begins decreasing with increasing $n_h$ -- signalling the onset of the regime where the charge carriers are electrons, expected in the paramagnetic regime. We note that recent results on doped random (disordered) all-to-all connected $t-J$ models show qualitatively similar behavior~\cite{Shackleton_PRL_2021} but with a peak feature at $n_h \approx 0.25-0.3$.
\addtt{(We also note that in Ref.~\cite{Shackleton_PRL_2021}, $\gamma$ associated with the specific heat $C(T)$ i.e. $\gamma = \lim_{T\rightarrow 0} C(T)/T$ was studied as a function of $n_h$. This should be qualitatively similar to the metallic weight studied here, since both quantities are a reflection of the density of low-energy gapless modes, and hence the charge carriers.)}
We imagine that a high amount of frustration arising out of randomness and all-to-all character destroys any potential intermediate phase in such models, thereby leading to a sharp distinction between a hole-dominated and electron-dominated regime. It will be interesting to explore further the interplay of magnetism and metallicity in the intermediate phase, and to determine what aspects of the observed features of $\alpha$ are generic, and which are specific to the ITHM and whether wavefunction-based methods can be used to diagnose the nature of quasiparticles in this regime~\cite{kumar_quasiparticle_t_J_2022}.
\\
\\
\noindent
{\bf{Impact on the finite $U/t$ Hubbard model}}\\ 
We now consider the case of finite, but large, $U/t$, at small $n_h$ with the intention of connecting to realistic situations, and to determine the crossover scale between HS antiferromagnetism and superexchange physics. We simulate the \addt{usual} onsite Hubbard model, \addt{as in Eq.~\eqref{eq:Hubbard}}, directly for this purpose, allowing for double occupancy.
\addt{For $U/t \gtrsim 10.6$} the ground state is known to be a $120^\circ$ AFM ground state at exactly half filling~\cite{Zaletel_SpinLiquid} -- this state survives for a small, but finite, density of holes; as has been reported by previous work~\cite{Li2014entropy}. \addt{(At half-filling, a metallic paramagnet exists for $U/t \lesssim 8.5$ and in the region $ 8.5 \lesssim U/t \lesssim 10.6$ a chiral spin liquid has been argued to exist~\cite{Zaletel_SpinLiquid}).} As $U/t$ grows larger, the effective \addt{superexchange scale} $J$ decreases, and eventually only the kinetic mechanism is expected to be important. The natural question that arises is -- what impact does kinetic frustration have on the observed finite $U$ behavior? 

\begin{figure}
        \includegraphics[clip, trim=0cm 0.42cm 0cm 0.22cm,width=\linewidth]{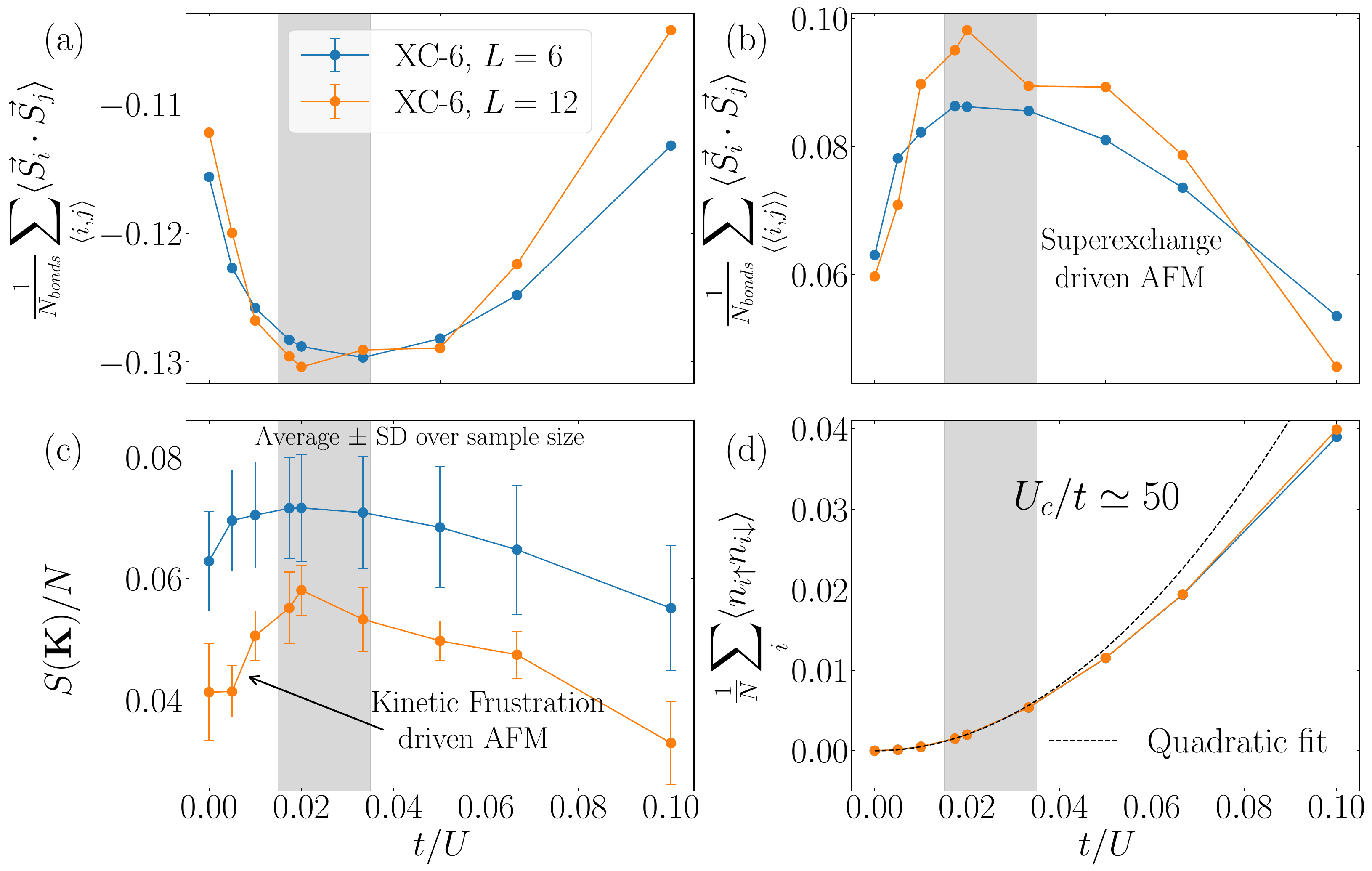}
        \caption{\justifying \textbf{\addtt{Interplay between kinetic magnetism and superexchange at low doping.}}
         \addtt{(a-d)} Ground state observables for $n_{h} \approx 0.11$ (\addt{for which the \addtt{$U/t=\infty$} ground state lies} within the \addtt{Haerter-Shastry antiferromagnetic phase}) as a function of $t/U$. (a) average spin correlation for nearest neighbor bonds of the triangular lattice, (b) average spin correlation for next-to-nearest neighbor bonds, (c) spin structure factor \addtt{$S(\vec{q}) = \sum_{j} e^{i \vec{q} \cdot (\vec{r}_\chi -\vec{r}_j)} \langle \vec{S_{\chi}} \cdot \vec{S_j} \rangle$ computed for reference site $\chi$, averaged over multiple $\chi$, at ordering vector $\vec{q}=\mathbf{K} = (4\pi/3,0)$. Error bars are obtained from the standard deviation (SD) of $S(\vec{q})$ for several reference sites chosen near the center. The number of reference sites -- our sample size for obtaining the average value and SD -- is 7 for XC-6, $L=6$ and 10 for XC-6, $L=12$} (d) average onsite double occupancy \addt{compared to a quadratic fit.}}
        \label{fig:fig4}
\end{figure}

At large $U/t$ and small $n_h$, there is a competition between spins and holes.   \addt{When double occupancy is negligible, the two terms in Eq.~\eqref{eq:tJmodel} compete.}
On the one hand, the system lowers its \addt{total} energy by delocalizing holes (maximizing \addt{the magnitude of} the kinetic energy), which comes at the cost of additional quantum fluctuations to the spin background, which in turn causes an effective reduction of local spin moments and hence a smaller contribution to the magnetic term. For finite $n_h$, both these energetic costs are macroscopically large. The associated loss of magnetic energy is \textit{not too unfavorable} when $J$ is small or zero. \addt{Thus, in this parameter} regime the holes dominate, i.e., the spins adjust to provide the most conducive environment for the holes to lower their \addt{total} energy. When $J$ increases (i.e., $U$ decreases, $t/U$ increases), the spins resist this movement of holes; correspondingly, the nearest neighbor AF spin correlations should increase in magnitude (\addt{i.e.} become more negative \addt{to lower the overall energy}). \addt{We refer to this regime as ``kinetic frustration driven AFM" in Fig.~\ref{fig:fig4}}.

Our intuitive \addt{and qualitative} picture is borne out in the \addt{quantitatively accurate} DMRG calculations, see Fig. \ref{fig:fig4}\textcolor{blue}{(a-c)}. \addt{
We have shown two system sizes at similar hole densities \addt{for which the $U/t = \infty$ ground state lies} in the HS regime.
We plot the expectation value of nearest neighbor (NN) and next-nearest neighbor (NNN)  spin-spin correlations in the ground state, as a function of $t/U$. We have averaged over all pairs of sites that have the same distance between them in real space (separately for the NN and NNN cases). This averaging suppresses anisotropies associated with the fact that the cylinder breaks translational symmetry in one direction, and the six-fold rotational symmetry of the triangular lattice.
We also plot the value of the structure factor at the ${\bf K}$ point, $S(\mathbf{K})$, expected to be large for the state with $120^\circ$ AFM order. This calculation is done for several central reference sites (each with method 2) and averaged. 
Error bars are obtained from the standard deviation, reflecting the degree of inhomogeneity of the state that was obtained in the DMRG.
We find that on starting from $t/U \rightarrow 0$, the antiferromagnetic NN (Fig.~\ref{fig:fig4}\textcolor{blue}{(a)}) and ferromagnetic NNN (Fig.~\ref{fig:fig4}\textcolor{blue}{(b)}) correlations increase in magnitude with increasing $t/U$, up to a certain scale that we will soon associate with a crossover.
$S(\mathbf{K})$ shown in Fig.~\ref{fig:fig4}\textcolor{blue}{(c)}, exhibits similar qualitative behavior as the local correlation functions.}

To understand the trends \addt{observed numerically} as \addt{$t/U$ is increased
further}, consider the situation 
from the lens of much lower (but still appreciably large) $U/t$. 
\addt{In this parameter regime the ground state has $120^{\circ}$ 
AFM order, but now due to superexchange.
For these values of $U/t$,} the $t-J$ model description is insufficient, since double occupancy is appreciable \addt{and cannot be ignored}; this is \addt{also} strongly tied to the fact that there are beyond-Heisenberg terms and these tend to destroy spin ordering tendencies~\cite{beyond_hesinberg2017,Zaletel_SpinLiquid}. As $t/U$ decreases, \addt{NN} AFM correlations strengthen because of the suppression of these effective interactions and the reduction of double occupancy, leading to the formation of stronger spin moments. Thus, spin moments strengthen with lowering $U/t$ starting from the HS limit, and they \addt{also} strengthen when increasing $U/t$ \addt{in the superexchange driven regime}, due to suppression of non-Heisenberg terms. \addt{This} must lead to a maximization of spin ordering tendencies at some $U/t$. We identify this as a crossover scale $U_c/t$. 

\addt{In Fig.~\ref{fig:fig4}\textcolor{blue}{(a-c)} we find this scale to be $U_c/t \approx 50$ for $n_h \approx 0.11$ 
and we have 
marked the location of the feature in grey.} Not surprisingly, the feature is rather broad and thus our inferred value must be taken to be only a ballpark estimate of this scale. \addt{We have also carried out additional calculations for other values of $n_h$ whose $U/t = \infty $ ground state lies in the HS regime, and found a similar crossover scale, (\addtt{see Methods -- Hole-density dependent crossover from the Haerter-Shastry to superexchange regime.}) The variation in the quantities plotted with $t/U$ becomes less prominent as $n_h$ is increased. Qualitatively, this may be anticipated since the role of $U$ is effectively diminished when there is more room for the particles to delocalize.} 

While \addt{the} discussion above focused on spin correlations, signatures of the crossover scale, \addt{though subtle}, exist even in observables associated with the charge degrees of freedom.
For example, local charge fluctuations are given by $\langle n_i^2 \rangle - \langle n_i \rangle^2 = \langle n_i \rangle (1 - \langle n_i \rangle) + 2 \langle n_{i,\uparrow}n_{i,\downarrow}\rangle$. As can be seen from the expression, these charge fluctuations are a direct measure of double occupancy, which we plot as a function of $t/U$ in Fig.~\ref{fig:fig4}\textcolor{blue}{(d)}. \addt{In the small $t/U$ regime, perturbative arguments suggest a quadratic dependence of this correlator on $t/U$. We find evidence for this numerically by viewing the same curve on a log-log plot (not shown). 
The dotted curve shown \addtt{corresponds} to a quadratic fit to our numerical data for small $t/U$. 
Then for $t/U \gtrsim 0.02$, i.e. $U/t \lesssim 50$, we observe a deviation from this quadratic dependence, which is consistent with the breakdown of the plain vanialla $t-J$ approach --  we thus associate this $U/t$ with the crossover scale.}

\addt{The scale of $U/t \sim 50$ for the case of low hole doping is large when compared to estimates for other strongly correlated materials, such as the cuprates~\cite{Future_SC_2025, arovas2022hubbard}. However, we note that this parameter regime is extremely relevant from the perspective of existing transition metal dichalcogenide bilayer materials. For example, Ref.~\cite{tao2024observation} experimentally studied bilayer MoTe$_2$/WSe$_2$ and estimated $t \approx 5$ meV, $J \approx 0.25$ meV. This translates to $t/J \approx 20$ and $U/t \approx 80$, which places the system in the kinetically driven AFM regime. Although the focus of Ref.~\cite{tao2024observation} was on spin polarons in a magnetic field, it will be interesting to determine the nature of its magnetism and transport in the zero field case.}
\addtt{We also note that this scale of $U/t$ lies in the range of what has been realized in cold-atom based platforms~\cite{lebrat2024observation}, thus our results should be testable in the near future}.
\\
\\
\noindent
\MakeUppercase{\textbf{Conclusions}}\\
In summary, we have studied ground-state magnetic and metallic properties of the infinite-$U$ triangular Hubbard model (ITHM) at finite hole density, and determined its impact on the case of realistic and finite $U/t$. Our investigation was motivated by considerations of Haerter-Shastry (HS) antiferromagnetism at $U = \infty$, arising from kinetic frustration, and its interplay with superexchange physics. We have identified an intermediate phase that is flanked by the HS 120$^\circ$ AFM on one side, and by the paramagnet on the other. This phase is a result of the holes relieving frustration -- this results in the formation of short-range dimers or more general multimers (multiply correlated spins), which also tend to organize themselves in stripe-like patterns. We also showed that the entire ITHM phase diagram (across all hole densities) shows signs of metallic behavior, and we used the metallic weight to observe the crossover from a hole-dominated regime to an electron-dominated regime. Our investigations focused on the case of zero temperature; however, it remains to be seen how the heavy Fermi liquid at low density, associated with the HS phase, potentially crosses over to a non-Fermi liquid ($T$-linear) regime that has been recently argued to exist on the square lattice~\cite{Fratini_2025}. 

At low hole density, we have identified a crossover scale between the HS and the superexchange-driven AFM as $U/t$ is tuned. 
Unlike the case of the square lattice, the ground state stabilized on the triangular lattice by both superexchange and kinetic frustration is qualitatively the same. We developed a physical picture for their competing tendencies and associated crossover, supported by DMRG calculations, and we have also provided several concrete predictions that should be testable in future experiments. Our findings are somewhat complementary to a recent theoretical proposal~\cite{Morera_2023} for explaining the 
increase in Curie Weiss temperature with hole doping in a triangular moiré material~\cite{tang}. 
In fact, Ref.~\cite{Morera_2023} has suggested a functional form for the magnetic susceptibility that modifies the Curie Weiss theory to have simple additive contributions from superexchange and HS antiferromagnetism, in turn suggesting that the two phenomena collaborate. Our current study shows the situation is subtler than such a form may suggest, and that further investigation is needed to understand the interplay of the two effects at finite $U/t$. 

Finally, our study focused only on onsite Hubbard interactions; it would be interesting to extend our work to the case of long-range or screened Coulomb interactions, as is common in many moiré materials~\cite{Xu2020}. Such interactions stabilize insulators e.g. charge-ordered states/generalized Wigner crystals at fractional fillings ~\cite{Xu2020,Li2021,kumar_wigner_2025}; the impact of kinetic frustration on the spin ordering of these phases~\cite{kim_kivelson_prb_2024} remains to be seen. Another important avenue would be to map out a thermal phase diagram to 
understand the interplay of kinetic frustration and superexchange at intermediate temperatures~\cite{lee2023triangular, Morera_2023}, a scale that is now readily accessible across multiple platforms~\cite{tang2020simulation, greinerXu}. Overall, we emphasize that while there has been a tremendous amount of effort devoted to the square lattice Hubbard model~\cite{Simons_Hubbard_2015, Future_SC_2025, arovas2022hubbard}, less is known about the frustrated cases -- our work shows that many surprises potentially lie in store in this area.
\\

\noindent
\MakeUppercase{\textbf{Methods}}\\
{\bf{Exact solution for one hole on small motifs}}\\
\\
{\bf{Single triangle}}\\
Consider the case of two up spins and one hole on a triangle -- the Hilbert space is three-dimensional. Define the three basis states as $|\uparrow \uparrow 0 \rangle \equiv c^{\dagger}_{1,\uparrow} c^{\dagger}_{2,\uparrow} | vac \rangle $, $|\uparrow 0 \uparrow \rangle \equiv c^{\dagger}_{1,\uparrow} c^{\dagger}_{3,\uparrow} | vac \rangle $ and $|0 \uparrow \uparrow \rangle \equiv c^{\dagger}_{2,\uparrow} c^{\dagger}_{3,\uparrow} | vac \rangle $. The two degenerate ground states, for $t>0$, with energy $E=-t$ are,
\begin{subequations}
\begin{eqnarray}
\frac{1}{\sqrt{3}}\Big( |\uparrow \uparrow 0 \rangle - \omega |\uparrow 0 \uparrow\rangle + \omega^2 |0 \uparrow \uparrow \rangle\Big)\\
\frac{1}{\sqrt{3}}\Big( |\uparrow \uparrow 0 \rangle - \omega^2 |\uparrow 0 \uparrow\rangle + \omega |0 \uparrow \uparrow \rangle\Big)
\end{eqnarray}
and the excited state with energy $E=2t$ is,
\begin{eqnarray}
\frac{1}{\sqrt{3}}\Big( |\uparrow \uparrow 0 \rangle - |\uparrow 0 \uparrow\rangle + |0 \uparrow \uparrow \rangle\Big)
\end{eqnarray}
\end{subequations}
Since the single triangle can be considered as a three-site ring, these states are simply momentum $2 \pi/3$, $-2 \pi/3$, and $0$ states, respectively. (Note that the relative minus sign accompanying the ket in the middle is a result of the fermionic normal ordering convention).

Next, consider the case of one up spin, one down spin, and one hole on a triangle -- the Hilbert space for the $t-J$ model (no double occupancy) is 6 dimensional. We find the ground state to be in the momentum zero sector, which is two-dimensional. The symmetrized states in this sector are,
\begin{subequations}
\begin{eqnarray}
|a \rangle & \equiv & \frac{1}{\sqrt{3}}\Big(|\uparrow \downarrow 0 \rangle + |0 \uparrow \downarrow\rangle - |\downarrow 0 \uparrow\rangle \Big) \\
|b \rangle & \equiv & \frac{1}{\sqrt{3}}\Big(|\downarrow \uparrow 0 \rangle + |0 \downarrow \uparrow\rangle - |\uparrow 0 \downarrow\rangle \Big). 
\end{eqnarray}
\end{subequations}
The Hamiltonian in this basis is,
\begin{eqnarray}
H = \begin{pmatrix}
0 & 2t \\
2t & 0   
\end{pmatrix}
\end{eqnarray}
The lowest energy state, for $t>0$, has energy $E=-2t$. The corresponding wavefunction is $\frac{1}{\sqrt{2}}\Big( |a\rangle - |b \rangle \Big)$, which in terms of the basis states is, 
\begin{align}
\frac{1}{\sqrt{6}} \Big( |\uparrow \downarrow 0 \rangle + |0 \uparrow \downarrow\rangle - |\downarrow 0 \uparrow\rangle - |\downarrow \uparrow 0 \rangle - |0 \downarrow \uparrow\rangle + |\uparrow 0 \downarrow\rangle \Big) 
\end{align}

\noindent
{\bf{$2\times2$ square unit cell}}\\
Consider the case of two up spins, one down spin, and one hole on the $2\times2$ square plaquette. In units of the lattice constant, we take site 1 to be located at (0,0), site 2 at (0,1), site 3 at (1,1) and site 4 at (1,0). The single-body translation operators are such that $T_x $ takes $1 \rightarrow 4, 2 \rightarrow 3, 3 \rightarrow 4, 4 \rightarrow 1$ and $T_y$ takes $1 \rightarrow 2, 2 \rightarrow 1, 3 \rightarrow 4, 4 \rightarrow 3$.  

For the $t-J$ model, the Hilbert space for this finite cluster is 12-dimensional. The convention for normal ordering follows the same scheme as earlier, for example, $|\uparrow \downarrow \uparrow 0 \rangle \equiv c^{\dagger}_{1,\uparrow}c^{\dagger}_{2,\downarrow}c^{\dagger}_{3,\uparrow} |vac \rangle$ and $|\downarrow 0 \uparrow \uparrow \rangle \equiv c^{\dagger}_{1,\downarrow}c^{\dagger}_{3,\uparrow}c^{\dagger}_{4,\uparrow} |vac \rangle$. 
We use translational symmetry to make the 12 dimensional Hamiltonian block diagonal, with each of the 4 blocks being 3-dimensional. The ground state is in the $(\pi,\pi)$ sector. In this sector, we write the three momentum-adapted states as,

\begin{subequations}
\begin{align}
|a \rangle &=& \frac{1}{2}\Big(| \uparrow \uparrow \downarrow 0 \rangle + |\uparrow \uparrow 0 \downarrow \rangle + |0 \downarrow \uparrow \uparrow \rangle + |\downarrow 0 \uparrow \uparrow \rangle \Big)\\
|b \rangle &=& \frac{1}{2}\Big(| \uparrow \downarrow \uparrow 0 \rangle + |\downarrow \uparrow 0 \uparrow \rangle + | 0 \uparrow \downarrow \uparrow \rangle + | \uparrow 0 \uparrow \downarrow \rangle \Big) \\
|c \rangle &=& \frac{1}{2}\Big(| \downarrow \uparrow \uparrow 0 \rangle + |\uparrow \downarrow 0 \uparrow \rangle + | 0 \uparrow \uparrow \downarrow \rangle + |\uparrow 0 \downarrow \uparrow \rangle \Big) 
\end{align}
\end{subequations}
The order of kets in a given momentum-adapted state corresponds to the action of $T_x^i T_y^j$ on the first ket and in the order $(i,j) = (0,0), (0,1), (1,0), (1,1)$ and the relative signs between kets emerge from both normal ordering and spatial translations (phase factors from momenta).
The matrix form (using ordering $|a\rangle, |b \rangle, |c \rangle$) of the $t-J$ Hamiltonian (with $J=0$) in this 3-dimensional Hilbert space is,
\begin{eqnarray}
H = -t \begin{pmatrix}
+1 & +1 & 0 \\
+1 & 0 & +1 \\
0 & +1 & +1 
\end{pmatrix}
\end{eqnarray}
For $t>0$, the ground state has energy $-2t$ and the corresponding state is
\begin{equation}
|\psi_{GS}\rangle = \frac{1}{\sqrt{3}}\Big( |a\rangle + |b\rangle + |c \rangle \Big)
\label{eq:psiGS_square}
\end{equation}
which is nothing but $\frac{1}{\sqrt{12}} \sum_i |B_i \rangle$ where $|B_i\rangle$ refer to all allowed basis states, for two up spins and one down spin. 

This result should not come across as a total surprise -- the Nagaoka condition is satisfied for a square plaquette and the fully polarized ferromagnet is
\begin{equation}
|FM \rangle = \frac{1}{2}\Big(|\uparrow \uparrow \uparrow 0 \rangle + |\uparrow \uparrow 0 \uparrow \rangle + | \uparrow 0 \uparrow \uparrow \rangle + |0 \uparrow\uparrow\uparrow\rangle \Big)
\end{equation}
which lies in the $(\pi,\pi)$ sector and has energy $-2t$. Applying $S^{-}_{tot} \equiv S^{-}_1 + S^{-}_2 + S^{-}_3 + S^{-}_4$ $|FM\rangle$ gives the spin-lowered multiplet, the state in Eq.~\eqref{eq:psiGS_square}, with exactly the same energy as $|FM\rangle$, a consequence of SU(2) symmetry of the model.\\

\noindent
{\bf{$2\times2$ triangular unit cell}}\\
Consider the case of two up spins, one down spin, and one hole on the $2\times2$ triangular unit cell, with exactly the same numbering as Fig.~\ref{fig:fig1}\textcolor{blue}{(b)}. The unit cell has not been repeated, but this is implicitly assumed. In this geometry, all four sites are connected by hopping, and see each other twice when periodic boundary conditions are incorporated. (We count the hoppings between any two sites only once, analogous to what we did in the square case). This particular configuration is of qualitative relevance to the case of $n_h=1/4$, and illuminates the origin of our findings from DMRG on much bigger system sizes.

We use translational symmetry, defining ``x" to be the horizontal direction, and ``y" to be the oblique direction at 60$^{\circ}$ with respect to the horizontal. The single-body translation operators are such that $T_x $ takes $1 \rightarrow 3, 2 \rightarrow 4, 3 \rightarrow 1, 4 \rightarrow 2$ and $T_y$ takes $1 \rightarrow 4, 2 \rightarrow 3, 3 \rightarrow 2, 4 \rightarrow 1$. Just like the square-case, the 12 dimensional Hamiltonian is block diagonalized by momentum labels $(k_x,k_y)$ with each $k_i=0$ or $\pi$, with each of the 4 blocks being 3-dimensional. We find that there is one ground state in each of the $(0,\pi), (\pi,0)$ and $(\pi,\pi)$ sectors, and none in $(0,0)$. 
 
In the $(\pi,\pi)$ sector we write the three momentum-adapted states, 
\begin{subequations}
\begin{align}
|a \rangle &=& \frac{1}{2}\Big(| \uparrow \uparrow \downarrow 0 \rangle + |0 \downarrow \uparrow \uparrow \rangle - | \downarrow 0 \uparrow \uparrow \rangle - |\uparrow \uparrow 0 \downarrow \rangle \Big)\\
|b \rangle &=& \frac{1}{2}\Big(| \uparrow \downarrow \uparrow 0 \rangle + |0 \uparrow \downarrow \uparrow \rangle - | \uparrow 0 \uparrow \downarrow \rangle - |\downarrow \uparrow 0 \uparrow \rangle \Big) \\
|c \rangle &=& \frac{1}{2}\Big(| \downarrow \uparrow \uparrow 0 \rangle + |0 \uparrow \uparrow \downarrow \rangle - | \uparrow 0 \downarrow \uparrow \rangle - |\uparrow \downarrow 0 \uparrow \rangle \Big) 
\end{align}
\end{subequations}
The matrix form (using ordering $|a\rangle, |b \rangle, |c \rangle$) of the $t-J$ Hamiltonian (with $J=0$) in this 3-dimensional Hilbert space is,
\begin{eqnarray}
H = -t \begin{pmatrix}
-1 & +1 & +1 \\
+1 & +1 & -1 \\
+1 & -1 & +1 
\end{pmatrix}
\end{eqnarray}
We find that the ground state, for $t>0$, has energy $-2t$ and the corresponding eigenvector is given by,
\begin{equation}
 |\psi_{gs} \rangle_{(\pi,\pi)} \equiv \frac{1}{\sqrt{2}}\Big( |b \rangle - |c \rangle \Big)
\end{equation}
which can be explicitly written out as,
\begin{equation}
\begin{split}
|\psi_{gs} \rangle_{(\pi,\pi)}  = \frac{\Big(|\uparrow \downarrow \rangle - |\downarrow \uparrow \rangle \Big)}{\sqrt{2}} \otimes \frac{\Big(|\uparrow 0 \rangle + |0 \uparrow \rangle \Big)}{\sqrt{2}} \\
-\frac{\Big(|\uparrow 0 \rangle + |0 \uparrow \rangle \Big)}{\sqrt{2}}
\otimes
\frac{\Big(|\uparrow \downarrow \rangle - |\downarrow \uparrow \rangle \Big)}{\sqrt{2}}
\end{split}
\end{equation}
This form is particularly illuminating -- it shows that the wavefunction is not a product state, but is a simple equal amplitude superposition of two such states. It indicates that with equal probability, opposite spins on 1 and 2 form a singlet while the other up spin is delocalized on the other two sites (3 and 4), or, alternatively, opposite spins on sites 3 and 4 form a singlet while the up spin is delocalized on sites 1 and 2.
We note that the expectation value of $\langle \vec{S}_1 \cdot \vec{S}_2 \rangle$ and $\langle \vec{S}_3 \cdot \vec{S}_4 \rangle $ are the only non zero ones, and equal half the value achieved in a perfect singlet, i.e. $-3/8$, while for every other $i \neq j$ $\langle \vec{S}_i \cdot \vec{S}_j \rangle = 0$. This is precisely the pattern of spin correlations reported in Fig.~\ref{fig:fig1}\textcolor{blue}{(b)} for one of the solutions.

Since the Hamiltonian for a $2\times2$ triangular cell is an all-to-all one, 
and all pairs are equivalent, the other 
two solutions (in other momentum sectors) correspond to the two 
other ways of ``dimer covering" the triangular unit cell. We have shown one of these coverings (1 with 4, 2 with 3) and have not shown the covering; 1 with 3, 2 with 4. We clarify that an appropriate linear combination of solutions yields the diamond pattern for spin correlations in Fig.~\ref{fig:fig1}\textcolor{blue}{(b)}.\\

\noindent
{\bf{Analysis of geometry and finite-size dependence of numerical results}}\\

\begin{figure}[]
    \centering
        \includegraphics[clip, trim=0cm 0cm 0cm 0cm,width=0.9\linewidth]{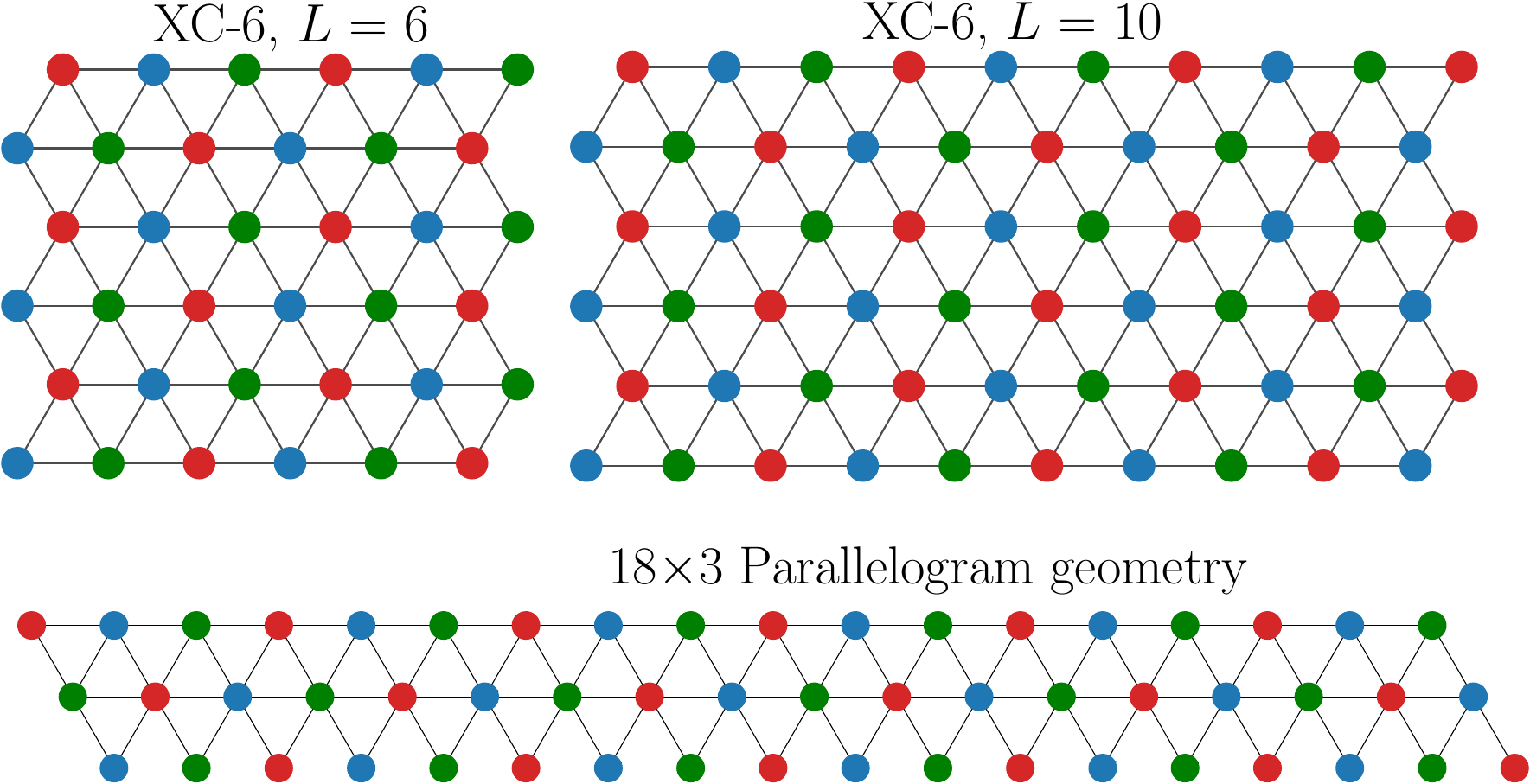}
    \caption{\justifying \textbf{\addtt{Examples of XC and parallelogram cylindrical samples of the type used in this study.}} \addtt{Periodic boundary conditions are applied along one direction and open boundary conditions are applied in the horizontal $x$- direction.} \addt{The top two correspond to the XC-6, length 6 and 10 cylinders, and the bottom one corresponds to a $18 \times 3$ parallelogram geometry.} The length 10 XC-6 cylinder does not support the $120^{\circ}$ AFM order, and hence does not accommodate the {$\bf K$} and symmetry related points.}
    \label{fig:cylinders}
\end{figure}

\begin{figure}
    \centering
    \includegraphics[width=1\linewidth]{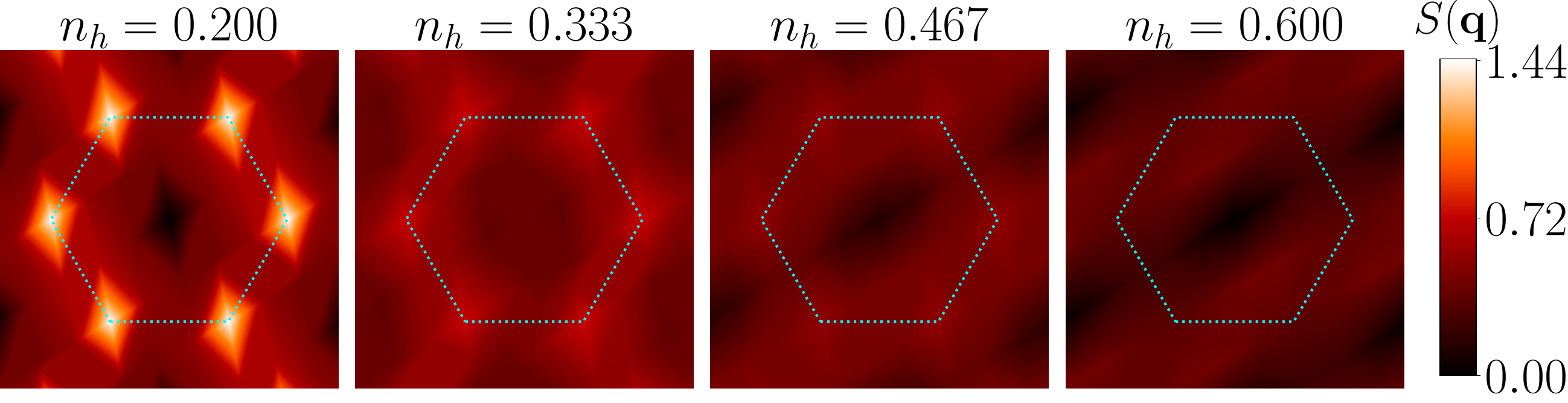}
    \caption{\textbf{\addtt{Spin structure factor from exact diagonalization.}} Ground state static spin structure factor $S(\vec{q})$, \addtt{as in Eq.\eqref{eq:spinstrfact}}, for \addtt{$U/t=\infty$} for the 15-site triangular geometry, computed with ED, for representative hole \addt{densities}, $n_{h}$.  The horizontal and vertical axes correspond to $q_x$ and $q_y$ respectively, and the center corresponds to
$(0,0)$. The hexagon marks the boundary of the first Brillouin zone of the triangular lattice.}
    \label{fig:15site}
\end{figure}

As mentioned in the main text, we have carried out DMRG calculations on multiple cylindrical geometries (see Fig. \ref{fig:cylinders} for some examples), including both XC and parallelogram geometries, in an effort to generalize our conclusions to the thermodynamic limit. To supplement our DMRG results, we have also carried out exact diagonalization (ED) for the ITHM for a wide range of $n_h$ on a 15-site cluster, which was previously studied in Ref.~\cite{lee2023triangular}. Our results for the hole density dependence of the spin structure factor are consistent with the DMRG results and are shown in Fig.~\ref{fig:15site}.

\begin{figure}[] 
    \centering
        \includegraphics[clip, trim=0cm 0cm 0cm 0cm,width=\linewidth]{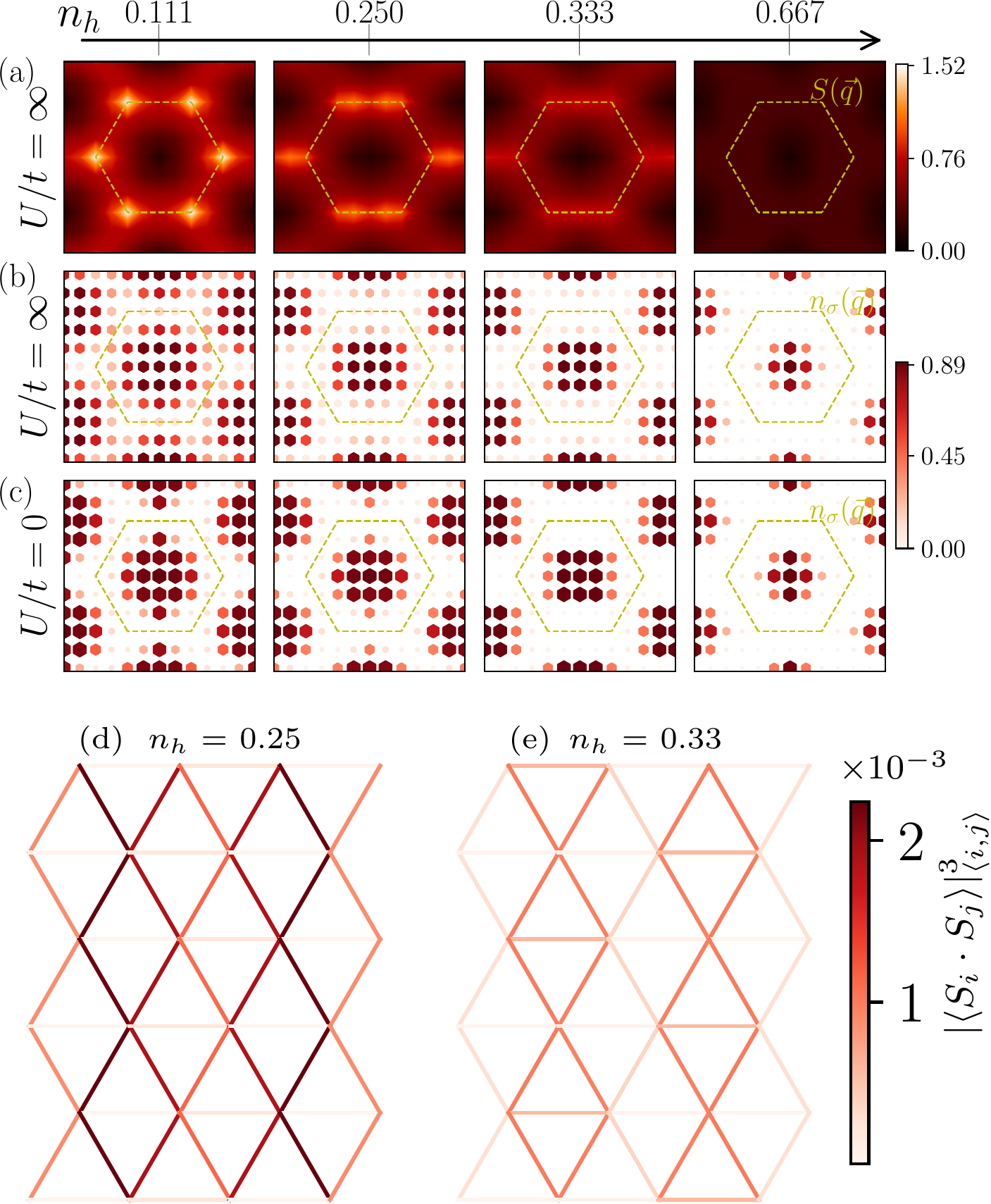}
    \caption{\justifying \textbf{\addtt{Magnetic and metallic properties induced by hole doping the triangular Hubbard model on the length 6 XC-6 cylinder}} (a) Ground state static spin structure factor $S(\vec{q})$, \addtt{as in Eq.\eqref{eq:spinstrfact}}, for the 36-site \addt{(XC-6, length 6 cylinder, restricted to sites in the bulk)} for $U/t=\infty$ for representative hole \addt{densities}, $n_{h}$. (b-c) Ground state momentum space fermionic occupation number of one spin species $n_{\sigma}(\vec{q})$, \addtt{as in Eq.\eqref{eq:mom_occ}}, for (b) $U/t = \infty$ and (c) $U/t = 0$. (d-e) Real space profile of spin correlations $\langle \vec{S_i}\cdot \vec{S_j}\rangle$ at (d) $n_h=1/4$ and (e) $n_h=1/3$. The red color variation on the bonds represents the strength of nearest-neighbor AFM correlations.}
    \label{fig:6x6}
\end{figure}

We show DMRG results for the \addt{length 6} XC-6 cluster in Fig.~\ref{fig:6x6}  and find resemblance with bigger system sizes (compare with Fig.~\ref{fig:fig2} of the main text). The HS phase
is stable up to $n_{h}\simeq0.2$, and the paramagnetic phase prominently appears above $n_{h}\simeq0.5$. The intermediate phase is associated with the breaking of the six-fold rotational symmetry in $S(\vec{q})$ (see panel a), and more directly, 
in the real space spin-spin correlations which show the formation of stripes and ``diamond" configuration. We have not established the precise order parameter of the intermediate phase, and whether it is related to the paramagnetic phase via a phase transition or crossover.

In the main text, we also remarked that the intermediate phase displays signatures of the physics of $n_h = 1/4$ and $n_h=1/3$. In this low hole density regime, we anticipated close competition between different kinds of local spin correlations, due to the three-fold degenerate ground states seen for the one hole problem on the $2 \times 2$ triangular unit cell. 

\begin{figure}[] 
    \centering
        \includegraphics[clip, trim=2.1cm 2.6cm 0.2cm 0.8cm,scale=0.5]{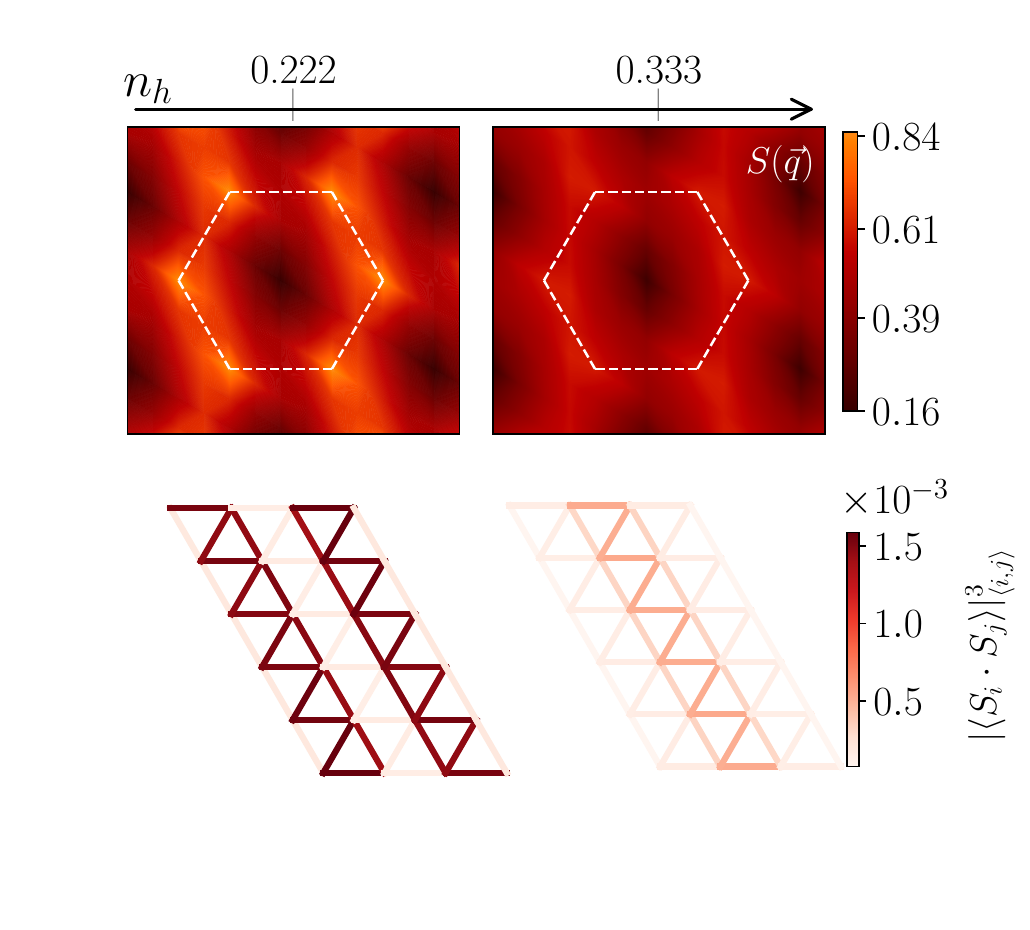}
    \caption{\justifying \textbf{\addtt{Magnetic properties induced by hole doping the triangular Hubbard model on the length 6, width 6 parallelogram geometry.}} DMRG results for $U/t=\infty$ on the 36-site ($6 \times 6$ parallelogram) cylinder for two hole densities $n_{h} = 0.222$, which corresponds to $n_h \approx 0.25$ when the boundary sites are eliminated, and $n_{h} = 0.333$. (top) Ground state static spin structure factor $S(\vec{q})$, \addtt{as in Eq.\eqref{eq:spinstrfact}}. (bottom) Real space profile of spin-spin correlations on nearest neighbor bonds $\langle i,j\rangle$ -- the color is a measure of $\Big|\langle \vec{S_i}\cdot \vec{S_j}\rangle\Big|^3$.}
    \label{fig:diagonal6x6}
\end{figure}

\begin{figure}[] 
    \centering
        \includegraphics[clip, trim=0cm 0.41cm 0cm 0.28cm,width=\linewidth]{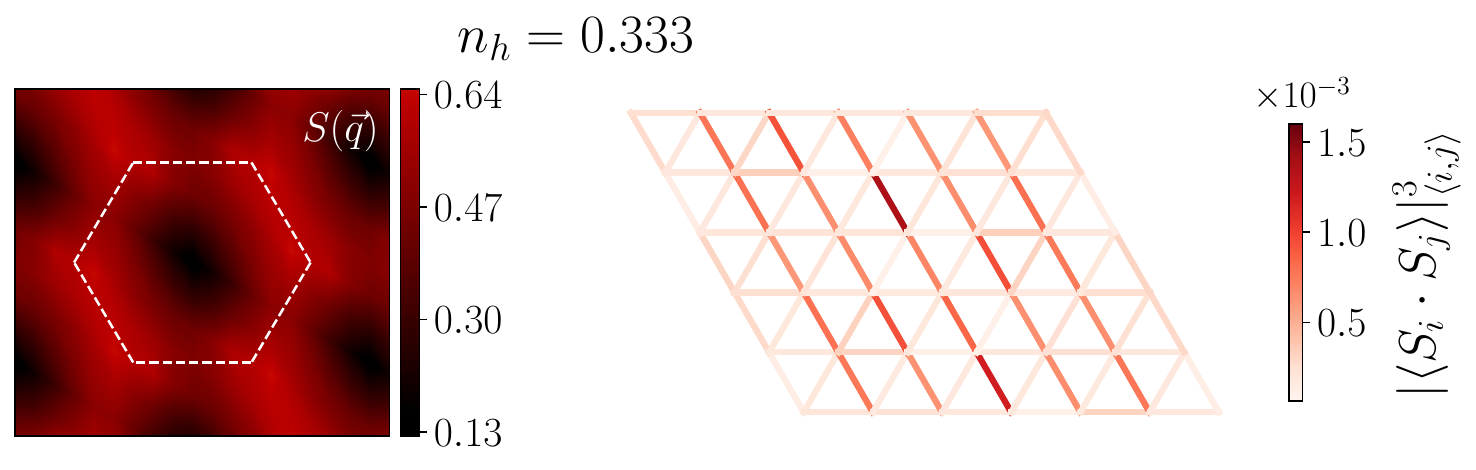}
 \caption{\justifying \textbf{\addtt{Magnetic properties induced by hole doping the triangular Hubbard model on the length 9, width 6 parallelogram geometry.}} DMRG results for $U/t=\infty$ on the 54-site ($9 \times 6$ parallelogram) cylinder for $n_{h} = 1/3$. The left and right panel descriptions are the same as those presented for Fig.~\ref{fig:diagonal6x6} (top and bottom).} 
   \label{fig:diagonal9x6}
\end{figure}

We show DMRG results on the $6 \times 6$ parallelogram geometry for $n_h=0.22$ and $n_h=0.33$ in Fig.~\ref{fig:diagonal6x6}, and on the $9 \times 6$ parallelogram geometry for $n_h=0.33$ in Fig.~\ref{fig:diagonal9x6}.
For both systems, we have eliminated boundary sites to highlight the nearest-neighbor spin-spin correlations in the center of the cluster. For the $6\times 6$ case we find that certain triangular motifs strengthen along the vertical direction giving the appearance of a ``triangular stripe", whereas in the latter case we find the formation of a different kind of stripe (``dimer" configuration). These results provide further evidence for the close competition between different kinds of local spin correlations for densities in the vicinity of $n_h=0.25$. We also plot the spin structure factor, which involves both short and long range correlations, and find that it does not show any sign of magnetic ordering for $n_h=0.33$ in both cases.\\

\noindent
{\bf{Analysis of antiferromagnetic ordering in the Haerter-Shastry regime}}\\
Previous work on the ITHM has shown that a single hole on top of the half-filled background favors 120$^\circ$ AFM order. It has also been argued that in the thermodynamic limit, the spin correlations should correspond to the classical value~\cite{sposetti}. In the absence of any symmetry breaking on a finite-size cluster, the ground state has definite $S^2$ -- stabilizing a classical symmetry-broken Neel state would require explicit symmetry breaking, as with application of a magnetic field, or the presence of degenerate ground states. In the thermodynamic limit, this occurs spontaneously since the quasidegenerate (Anderson) tower of states, each with definite $S^2$, can mix to yield a classical state~\cite{Bernu_1992,Bernu_1994}.

\begin{figure}[]
        \includegraphics[width=\linewidth]{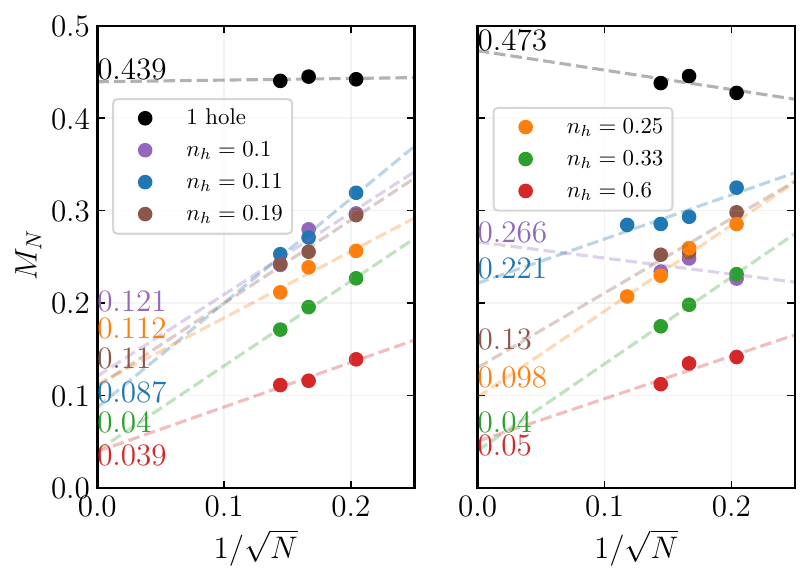}
    \caption{\justifying \textbf{\addtt{Finite-size scaling of the antiferromagnetic order parameter.}} Antiferromagnetic order parameter $M_{N}$ \addtt{$\equiv \sqrt{2S(\mathbf{K})/(N+6})$} versus $1/\sqrt{N}$ across a representative sample of hole densities $n_{h}$, extrapolated to the thermodynamic limit. We calculate $S(\mathbf{K})$ using \addtt{Eq.\eqref{eq:spinstrfact}} (left) and using
    \addtt{$S(\vec{q}) = \sum_{j} e^{i \vec{q} \cdot (\vec{r}_\chi -\vec{r}_j)} \langle \vec{S_{\chi}} \cdot \vec{S_j} \rangle$}
    for representative reference site $\chi$ near the center(right)} 
    \label{fig:magnetic order parameter}
\end{figure}

Following work by Ref.~\cite{Bernu_1992}, we examine the order parameter associated with AFM ordering on the triangular lattice on a finite-sized lattice of $N$ sites-- 
\begin{equation} 
M_{N} \equiv \sqrt{\frac{2S(\mathbf{K})}{N+6}}
\label{eq:mop}
\end{equation}
where $S(\mathbf{K})$ is the spin structure factor at the $\mathbf{K}$ and symmetry-related points, as defined in main text.  
Additionally, we extrapolate our results to the thermodynamic limit, assuming the functional form used for the Heisenberg AFM on the triangular lattice, 
\begin{equation}
M_{N} = M_{\infty} + \frac{a}{\sqrt{N}}.
\label{eq:mop_extrapolate}
\end{equation}
In either of the FM, paramagnetic, and short-range AFM states, $M_{\infty}$ should be strictly zero. $M_{\infty}$ for the classical 120$^{\circ}$ state is exactly equal to $1/2$. 
However, it must be kept in mind that the fitting form misses higher-order corrections in inverse system size, and must be thought of as only approximate in the range of sizes that we have studied here. It also assumes that the ground state is homogeneous, i.e. translationally invariant, and thus not strictly applicable for open cylinders. In cases where the extrapolation is sensitive to the method of choice, our analysis at best provides only a ballpark estimate of the order parameter. 

Fig.~\ref{fig:magnetic order parameter} shows that our DMRG results for the case of one hole are consistent with classical $120^{\circ}$ AFM ordering, though there is a small discrepancy with the expected value. For the case of low hole density, which on our system sizes corresponds to only a few holes, we find that are conclusions are influenced by geometry-dependent effects. First, as mentioned in main text, there is some tendency for particles to migrate towards the right or left edge of the cylinder and so the bulk density may not exactly correspond to $n_h$ -- the issue is of most concern \addt{for small} $n_h$. Closely tied to this issue is that the holes appear to localize easily when their density is low. For example, for $n_h \approx 0.11$, holes appear to be ``trapped" into domains, which makes the estimation of the order parameter less reliable compared to the situation when the system's response is homogeneous throughout the bulk, as is the case at $n_h = 0.25$. Taking the structure factor as defined with respect to spin-spin correlations involving only a single reference site (close to the center of the cylinder) ameliorates this situation to some extent -- while we are unable to definitively determine the precise value -- we conclude that the order parameter is non-zero.
The trends are consistent with the order parameter decreasing continuously with $n_h$, thus we expect a second-order quantum phase transition from the HS phase to the intermediate phase.
The order parameter for densities $n_h \gtrsim 0.3$ appears to be small and consistent with zero.

From the point of view of our numerical calculations, the selection of inhomogeneous states at \addt{small} $n_h$ is obviously undesirable, and primarily triggered by the short width direction, but it reveals some of the underlying physics of the problem that is responsible for these features. The low $Z$ factor means that the holes are heavy in the low-energy effective theory, equivalently, the band is flat. This close competition between multiple states may direct the DMRG optimization to choose states where holes tend to favor AFM textures in domains.\\

\noindent
{\bf{Determining quasiparticle weight}}\\
\begin{figure} 
        \includegraphics[clip, trim=0.25cm 0.25cm 0.2cm 0.2cm,scale=1]{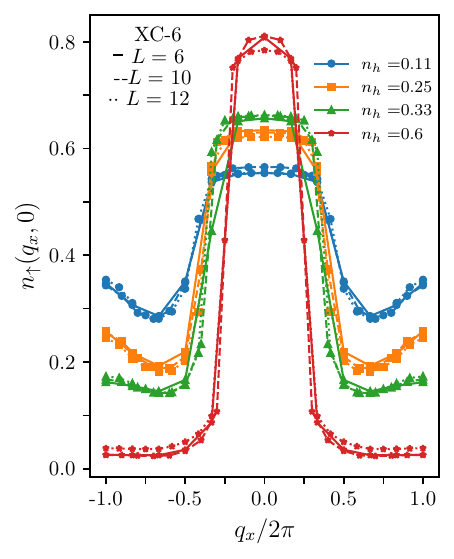}
    \caption{\justifying \textbf{\addtt{Momentum space occupation and quasiparticle weight.}} Ground state momentum space occupation number $n_{\uparrow}(q_{x},0)$, \addtt{as in Eq.\eqref{eq:mom_occ}}, versus $q_{x}/2\pi$ for \addtt{$U/t=\infty$} for representative XC-6 cylinders \addtt{of length $L$} and hole densities \addtt{$n_h$}. The quasiparticle weight was determined from the size of the jump, as explained further in the text.} 
     \label{fig:linecuts}
\end{figure}
In main text, we discussed how the quasiparticle weight (residue) $Z$ varies with hole density $n_h$. $Z$ was used to rescale the non-interacting ($U=0$) ground-state energy, and it was shown that the ground state energy (from DMRG) for the ITHM was explained by such a procedure. Here we show representative plots for how $Z$ was extracted from our DMRG ground state wavefunctions.

Fig.~\ref{fig:linecuts} shows the occupation number $n_{\sigma}(\vec{q})$ computed in the quantum ground state, for a particular spin species $\sigma = \uparrow$, as a function of one-dimensional momentum $q_x$, keeping $q_y=0$ fixed. Results for three representative system sizes and four representative hole densities are plotted. The agreement across system sizes suggests that the observed trends and conclusions drawn in the main text persist in the thermodynamic limit. We obtained the Fermi momentum ($\vec{q}_{F}$) for the non-interacting tight-binding system on exactly the same lattice, by plotting $n_{\sigma}(\vec{q})$ and determining the location of the jump. (We have ignored any  small errors introduced by open shell effects.)
\addt{This $q_F$ was used as a reference to identify the interacting one. However, due to the smearing out of the occupation for the interacting case, there is some uncertainty in determining its exact location which in turn leads to an uncertainty in $Z$. Therefore, the $Z$ values shown in the main text correspond to an estimate. The difference in the largest and smallest occupations provides upper \addtt{and lower} bounds on $Z$, this was used to obtain the uncertainty shown in Fig.~\ref{fig:fig3}(a) of the main text.}\\

\begin{figure}[H]
    \centering
    \includegraphics[width=\linewidth]{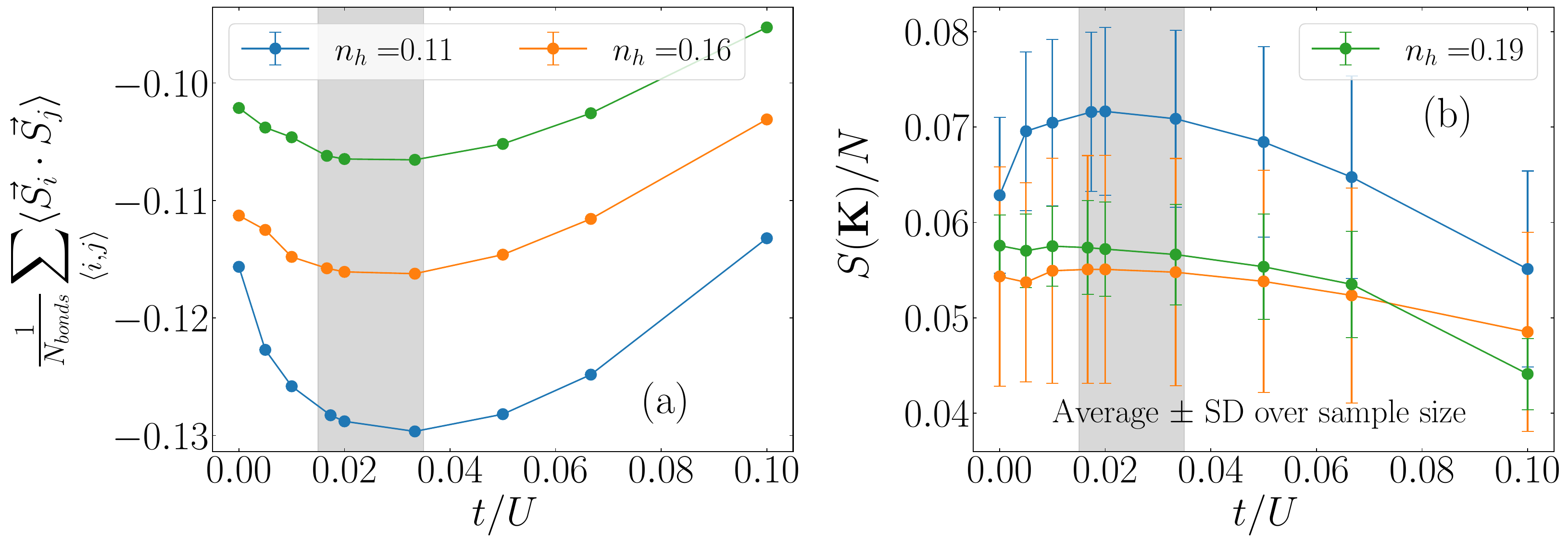}
    \caption{\justifying 
         \addt{\textbf{\addtt{Interplay between kinetic magnetism and superexchange across hole doping.}} \addtt{(a-b)} Ground state observables for the length 6 XC-6 cylinder for representative values of $n_{h}$ (whose $U/t = \infty$ ground state lies within the \addtt{Haerter-Shastry antiferromagnetic phase}) as a function of $t/U$. (a) average spin correlation for nearest neighbor bonds of the triangular lattice, (b) spin structure factor \addtt{$S(\vec{q}) = \sum_{j} e^{i \vec{q} \cdot (\vec{r}_\chi -\vec{r}_j)} \langle \vec{S_{\chi}} \cdot \vec{S_j} \rangle$ computed for reference site $\chi$, averaged over multiple $\chi$, at ordering vector $\vec{q}=\mathbf{K}=(4\pi/3,0)$. Error bars are obtained from the standard deviation (SD) of $S(\vec{q})$ for a sample size of 7 reference sites near the center.}}}
        
    \label{fig:crossover_at_diff_nh}
\end{figure}

\noindent
{\bf{Hole-density dependent crossover from the Haerter-Shastry to superexchange regime}}\\
\addt{In the main text we discussed the crossover between the kinetic AFM dominated regime and the superexchange regime for a representative value of $n_h \approx 0.11$.
In Fig.~\ref{fig:crossover_at_diff_nh} we show additional results for two additional representative $n_h$ for which the ground state at $U \rightarrow \infty$ lies in the HS regime. We compute the average nearest neighbor correlator, and the structure factor at the ${\bf K}$ point as a function of $t/U$ using procedures identical to those described in the main text. The error bars reflect the inhomogeneous state obtained \addtt{from DMRG}. We focus here on the general trends of their average value.}

\addt{We find that the crossover, characterized by the local maximum in the magnitude of these quantities, occurs over a similar range of $t/U$ throughout the HS regime. Since crossovers are broad, it is difficult to pinpoint an exact trend for the movement of this feature with increasing $n_h$. On the one hand, higher $n_h$ provides more channels for particles to avoid one another, which means that even low $U$ potentially mimics the kinetic effects of $U \rightarrow \infty$. On the other hand, the increase in hole density also promotes higher order exchange processes (which require finite $U$); these tend to suppress spin moment formation and thus push the crossover scale to higher $U$. The net result of these competing tendencies appears to minimally affect the ballpark in which the local maxima (of the magnitude) occur -- this could sensitively depend on the observable being examined.
Notwithstanding this subtlety, the overall magnitude of both quantities plotted decreases with increasing $n_h$, suggesting that AFM correlations typically weaken on increasing $n_h$. In particular, the difference between the maximum value of $S(\mathbf{K})$ and the $U=\infty$ result decreases with increasing $n_h$.}\\

\noindent
\MakeUppercase{\textbf{Data availability}}\\
The data as text files are openly available via
GitHub at \url{https://github.com/stsherif/Data_for_HS_ITHM_kinetic}. Requests for additional materials should be directed to S.S. and H.J.C.
\\

\noindent
\addtt{
\MakeUppercase{\textbf{Code availability}}\\
The post-processing python scripts used to generate the figures in the present study are openly available via
GitHub at \url{https://github.com/stsherif/Data_for_HS_ITHM_kinetic}. The DMRG and ED calculations were carried out using the open source ITensor~\cite{ITensor} and  
QuSpin libraries~\cite{quspin} respectively.}
\\

\noindent
\MakeUppercase{\textbf{References}}\\

\begin{thebibliography}{65}%
\makeatletter
\providecommand \@ifxundefined [1]{%
 \@ifx{#1\undefined}
}%
\providecommand \@ifnum [1]{%
 \ifnum #1\expandafter \@firstoftwo
 \else \expandafter \@secondoftwo
 \fi
}%
\providecommand \@ifx [1]{%
 \ifx #1\expandafter \@firstoftwo
 \else \expandafter \@secondoftwo
 \fi
}%
\providecommand \natexlab [1]{#1}%
\providecommand \enquote  [1]{``#1''}%
\providecommand \bibnamefont  [1]{#1}%
\providecommand \bibfnamefont [1]{#1}%
\providecommand \citenamefont [1]{#1}%
\providecommand \href@noop [0]{\@secondoftwo}%
\providecommand \href [0]{\begingroup \@sanitize@url \@href}%
\providecommand \@href[1]{\@@startlink{#1}\@@href}%
\providecommand \@@href[1]{\endgroup#1\@@endlink}%
\providecommand \@sanitize@url [0]{\catcode `\\12\catcode `\$12\catcode
  `\&12\catcode `\#12\catcode `\^12\catcode `\_12\catcode `\%12\relax}%
\providecommand \@@startlink[1]{}%
\providecommand \@@endlink[0]{}%
\providecommand \url  [0]{\begingroup\@sanitize@url \@url }%
\providecommand \@url [1]{\endgroup\@href {#1}{\urlprefix }}%
\providecommand \urlprefix  [0]{URL }%
\providecommand \Eprint [0]{\href }%
\providecommand \doibase [0]{https://doi.org/}%
\providecommand \selectlanguage [0]{\@gobble}%
\providecommand \bibinfo  [0]{\@secondoftwo}%
\providecommand \bibfield  [0]{\@secondoftwo}%
\providecommand \translation [1]{[#1]}%
\providecommand \BibitemOpen [0]{}%
\providecommand \bibitemStop [0]{}%
\providecommand \bibitemNoStop [0]{.\EOS\space}%
\providecommand \EOS [0]{\spacefactor3000\relax}%
\providecommand \BibitemShut  [1]{\csname bibitem#1\endcsname}%
\let\auto@bib@innerbib\@empty
\bibitem [{\citenamefont {Nagaoka}(1966)}]{NagaokaFM}%
  \BibitemOpen
  \bibfield  {author} {\bibinfo {author} {\bibfnamefont {Y.}~\bibnamefont
  {Nagaoka}},\ }\bibfield  {title} {\bibinfo {title} {\textit{Ferromagnetism in
  a Narrow, Almost Half-Filled $s$ Band}},\ }\bibfield  {journal} {\bibinfo
  {journal} {Phys. Rev.}\ }\textbf {\bibinfo {volume} {147}},\ \href
  {https://doi.org/10.1103/PhysRev.147.392} {10.1103/PhysRev.147.392} (\bibinfo
  {year} {(1966)})\BibitemShut {NoStop}%
\bibitem [{\citenamefont {Thouless}(1965)}]{Thouless_1965}%
  \BibitemOpen
  \bibfield  {author} {\bibinfo {author} {\bibfnamefont {D.~J.}\ \bibnamefont
  {Thouless}},\ }\bibfield  {title} {\bibinfo {title} {\textit{Exchange in
  solid 3He and the Heisenberg Hamiltonian}},\ }\href
  {https://doi.org/10.1088/0370-1328/86/5/301} {\bibfield  {journal} {\bibinfo
  {journal} {Proceedings of the Physical Society}\ }\textbf {\bibinfo {volume}
  {86}},\ \bibinfo {pages} {893} (\bibinfo {year} {1965})}\BibitemShut
  {NoStop}%
\bibitem [{\citenamefont {Chao}\ \emph {et~al.}(1978)\citenamefont {Chao},
  \citenamefont {Spa\l{}ek},\ and\ \citenamefont {Ole\ifmmode~\acute{s}\else
  \'{s}\fi{}}}]{Spalek_1978}%
  \BibitemOpen
  \bibfield  {author} {\bibinfo {author} {\bibfnamefont {K.~A.}\ \bibnamefont
  {Chao}}, \bibinfo {author} {\bibfnamefont {J.}~\bibnamefont {Spa\l{}ek}},\
  and\ \bibinfo {author} {\bibfnamefont {A.~M.}\ \bibnamefont
  {Ole\ifmmode~\acute{s}\else \'{s}\fi{}}},\ }\bibfield  {title} {\bibinfo
  {title} {Canonical perturbation expansion of the hubbard model},\ }\href
  {https://doi.org/10.1103/PhysRevB.18.3453} {\bibfield  {journal} {\bibinfo
  {journal} {Phys. Rev. B}\ }\textbf {\bibinfo {volume} {18}},\ \bibinfo
  {pages} {3453} (\bibinfo {year} {1978})}\BibitemShut {NoStop}%
\bibitem [{\citenamefont {Hubbard}(1963)}]{Hubbard_1963}%
  \BibitemOpen
  \bibfield  {author} {\bibinfo {author} {\bibfnamefont {J.}~\bibnamefont
  {Hubbard}},\ }\bibfield  {title} {\bibinfo {title} {Electron correlations in
  narrow energy bands},\ }\href {https://doi.org/10.1098/rspa.1963.0204}
  {\bibfield  {journal} {\bibinfo  {journal} {Proceedings of the Royal Society
  of London. A. Mathematical and Physical Sciences}\ }\textbf {\bibinfo
  {volume} {276}},\ \bibinfo {pages} {238} (\bibinfo {year} {1963})},\ \Eprint
  {https://arxiv.org/abs/https://royalsocietypublishing.org/rspa/article-pdf/276/1365/238/54456/rspa.1963.0204.pdf}
  {https://royalsocietypublishing.org/rspa/article-pdf/276/1365/238/54456/rspa.1963.0204.pdf}
  \BibitemShut {NoStop}%
\bibitem [{\citenamefont {Tasaki}(1989)}]{Tasaki_1989}%
  \BibitemOpen
  \bibfield  {author} {\bibinfo {author} {\bibfnamefont {H.}~\bibnamefont
  {Tasaki}},\ }\bibfield  {title} {\bibinfo {title} {\textit{Extension of
  Nagaoka’s theorem on the large-U Hubbard model}},\ }\href
  {https://doi.org/10.1103/PhysRevB.40.9192} {\bibfield  {journal} {\bibinfo
  {journal} {Physical Review B}\ }\textbf {\bibinfo {volume} {40}},\ \bibinfo
  {pages} {9192} (\bibinfo {year} {1989})}\BibitemShut {NoStop}%
\bibitem [{\citenamefont {Roth}(1969)}]{Roth_1969}%
  \BibitemOpen
  \bibfield  {author} {\bibinfo {author} {\bibfnamefont {L.~M.}\ \bibnamefont
  {Roth}},\ }\bibfield  {title} {\bibinfo {title} {\textit{Electron Correlation
  in Narrow Energy Bands. II. One Reversed Spin in an Otherwise Fully Aligned
  Narrow $S$ Band}},\ }\href {https://doi.org/10.1103/PhysRev.186.428}
  {\bibfield  {journal} {\bibinfo  {journal} {Phys. Rev.}\ }\textbf {\bibinfo
  {volume} {186}},\ \bibinfo {pages} {428} (\bibinfo {year}
  {1969})}\BibitemShut {NoStop}%
\bibitem [{\citenamefont {Doucot}\ and\ \citenamefont
  {Wen}(1989)}]{Doucot_PRB_1989}%
  \BibitemOpen
  \bibfield  {author} {\bibinfo {author} {\bibfnamefont {B.}~\bibnamefont
  {Doucot}}\ and\ \bibinfo {author} {\bibfnamefont {X.~G.}\ \bibnamefont
  {Wen}},\ }\bibfield  {title} {\bibinfo {title} {\textit{Instability of the
  Nagaoka state with more than one hole}},\ }\href
  {https://doi.org/10.1103/PhysRevB.40.2719} {\bibfield  {journal} {\bibinfo
  {journal} {Phys. Rev. B}\ }\textbf {\bibinfo {volume} {40}},\ \bibinfo
  {pages} {2719} (\bibinfo {year} {1989})}\BibitemShut {NoStop}%
\bibitem [{\citenamefont {Shastry}\ \emph {et~al.}(1990)\citenamefont
  {Shastry}, \citenamefont {Krishnamurthy},\ and\ \citenamefont
  {Anderson}}]{ska}%
  \BibitemOpen
  \bibfield  {author} {\bibinfo {author} {\bibfnamefont {B.~S.}\ \bibnamefont
  {Shastry}}, \bibinfo {author} {\bibfnamefont {H.~R.}\ \bibnamefont
  {Krishnamurthy}},\ and\ \bibinfo {author} {\bibfnamefont {P.~W.}\
  \bibnamefont {Anderson}},\ }\bibfield  {title} {\bibinfo {title}
  {\textit{Instability of the Nagaoka ferromagnetic state of the
  $U=\ensuremath{\infty}$ \MakeUppercase{Hubbard} model}},\ }\bibfield
  {journal} {\bibinfo  {journal} {Phys. Rev. B}\ }\textbf {\bibinfo {volume}
  {41}},\ \href {https://doi.org/10.1103/PhysRevB.41.2375}
  {10.1103/PhysRevB.41.2375} (\bibinfo {year} {(1990)})\BibitemShut {NoStop}%
\bibitem [{\citenamefont {Basile}\ and\ \citenamefont
  {Elser}(1990)}]{Basile_Elser}%
  \BibitemOpen
  \bibfield  {author} {\bibinfo {author} {\bibfnamefont {A.~G.}\ \bibnamefont
  {Basile}}\ and\ \bibinfo {author} {\bibfnamefont {V.}~\bibnamefont {Elser}},\
  }\bibfield  {title} {\bibinfo {title} {\textit{Stability of the ferromagnetic
  state with respect to a single spin flip: Variational calculations for the
  U=\ensuremath{\infty} Hubbard model on the square lattice}},\ }\href
  {https://doi.org/10.1103/PhysRevB.41.4842} {\bibfield  {journal} {\bibinfo
  {journal} {Phys. Rev. B}\ }\textbf {\bibinfo {volume} {41}},\ \bibinfo
  {pages} {4842} (\bibinfo {year} {1990})}\BibitemShut {NoStop}%
\bibitem [{\citenamefont {Hanisch}\ \emph {et~al.}(1997)\citenamefont
  {Hanisch}, \citenamefont {Uhrig},\ and\ \citenamefont
  {M\"uller-Hartmann}}]{Hanisch_PRB_1997}%
  \BibitemOpen
  \bibfield  {author} {\bibinfo {author} {\bibfnamefont {T.}~\bibnamefont
  {Hanisch}}, \bibinfo {author} {\bibfnamefont {G.~S.}\ \bibnamefont {Uhrig}},\
  and\ \bibinfo {author} {\bibfnamefont {E.}~\bibnamefont
  {M\"uller-Hartmann}},\ }\bibfield  {title} {\bibinfo {title} {\textit{Lattice
  dependence of saturated ferromagnetism in the Hubbard model}},\ }\href
  {https://doi.org/10.1103/PhysRevB.56.13960} {\bibfield  {journal} {\bibinfo
  {journal} {Phys. Rev. B}\ }\textbf {\bibinfo {volume} {56}},\ \bibinfo
  {pages} {13960} (\bibinfo {year} {1997})}\BibitemShut {NoStop}%
\bibitem [{\citenamefont {Becca}\ and\ \citenamefont
  {Sorella}(2001)}]{Becca_PRL_2001}%
  \BibitemOpen
  \bibfield  {author} {\bibinfo {author} {\bibfnamefont {F.}~\bibnamefont
  {Becca}}\ and\ \bibinfo {author} {\bibfnamefont {S.}~\bibnamefont
  {Sorella}},\ }\bibfield  {title} {\bibinfo {title} {\textit{Nagaoka
  Ferromagnetism in the Two-Dimensional Infinite- $\mathit{U}$ Hubbard
  Model}},\ }\href {https://doi.org/10.1103/PhysRevLett.86.3396} {\bibfield
  {journal} {\bibinfo  {journal} {Phys. Rev. Lett.}\ }\textbf {\bibinfo
  {volume} {86}},\ \bibinfo {pages} {3396} (\bibinfo {year}
  {2001})}\BibitemShut {NoStop}%
\bibitem [{\citenamefont {Haerter}\ and\ \citenamefont
  {Shastry}(2005)}]{Haerter_PRL_2005}%
  \BibitemOpen
  \bibfield  {author} {\bibinfo {author} {\bibfnamefont {J.~O.}\ \bibnamefont
  {Haerter}}\ and\ \bibinfo {author} {\bibfnamefont {B.~S.}\ \bibnamefont
  {Shastry}},\ }\bibfield  {title} {\bibinfo {title} {\textit{Kinetic
  Antiferromagnetism in the Triangular Lattice}},\ }\bibfield  {journal}
  {\bibinfo  {journal} {Phys. Rev. Lett.}\ }\textbf {\bibinfo {volume} {95}},\
  \href {https://doi.org/10.1103/PhysRevLett.95.087202}
  {10.1103/PhysRevLett.95.087202} (\bibinfo {year} {(2005)})\BibitemShut
  {NoStop}%
\bibitem [{\citenamefont {Yin}\ \emph {et~al.}(2011)\citenamefont {Yin},
  \citenamefont {Haule},\ and\ \citenamefont {Kotliar}}]{Yin2011}%
  \BibitemOpen
  \bibfield  {author} {\bibinfo {author} {\bibfnamefont {Z.~P.}\ \bibnamefont
  {Yin}}, \bibinfo {author} {\bibfnamefont {K.}~\bibnamefont {Haule}},\ and\
  \bibinfo {author} {\bibfnamefont {G.}~\bibnamefont {Kotliar}},\ }\bibfield
  {title} {\bibinfo {title} {\textit{Kinetic frustration and the nature of the
  magnetic and paramagnetic states in iron pnictides and
  iron chalcogenides}},\ }\href {https://doi.org/10.1038/nmat3120} {\bibfield
  {journal} {\bibinfo  {journal} {Nature Materials}\ }\textbf {\bibinfo
  {volume} {10}},\ \bibinfo {pages} {932} (\bibinfo {year} {2011})}\BibitemShut
  {NoStop}%
\bibitem [{\citenamefont {Sposetti}\ \emph {et~al.}(2014)\citenamefont
  {Sposetti}, \citenamefont {Bravo}, \citenamefont {Trumper}, \citenamefont
  {Gazza},\ and\ \citenamefont {Manuel}}]{sposetti}%
  \BibitemOpen
  \bibfield  {author} {\bibinfo {author} {\bibfnamefont {C.~N.}\ \bibnamefont
  {Sposetti}}, \bibinfo {author} {\bibfnamefont {B.}~\bibnamefont {Bravo}},
  \bibinfo {author} {\bibfnamefont {A.~E.}\ \bibnamefont {Trumper}}, \bibinfo
  {author} {\bibfnamefont {C.~J.}\ \bibnamefont {Gazza}},\ and\ \bibinfo
  {author} {\bibfnamefont {L.~O.}\ \bibnamefont {Manuel}},\ }\bibfield  {title}
  {\bibinfo {title} {\textit{Classical Antiferromagnetism in Kinetically
  Frustrated Electronic Models}},\ }\bibfield  {journal} {\bibinfo  {journal}
  {Phys. Rev. Lett.}\ }\textbf {\bibinfo {volume} {112}},\ \href
  {https://doi.org/10.1103/PhysRevLett.112.187204}
  {10.1103/PhysRevLett.112.187204} (\bibinfo {year} {(2014)})\BibitemShut
  {NoStop}%
\bibitem [{\citenamefont {Lisandrini}\ \emph {et~al.}(2017)\citenamefont
  {Lisandrini}, \citenamefont {Bravo}, \citenamefont {Trumper}, \citenamefont
  {Manuel},\ and\ \citenamefont {Gazza}}]{lisandrini2017evolution}%
  \BibitemOpen
  \bibfield  {author} {\bibinfo {author} {\bibfnamefont {F.~T.}\ \bibnamefont
  {Lisandrini}}, \bibinfo {author} {\bibfnamefont {B.}~\bibnamefont {Bravo}},
  \bibinfo {author} {\bibfnamefont {A.~E.}\ \bibnamefont {Trumper}}, \bibinfo
  {author} {\bibfnamefont {L.~O.}\ \bibnamefont {Manuel}},\ and\ \bibinfo
  {author} {\bibfnamefont {C.~J.}\ \bibnamefont {Gazza}},\ }\bibfield  {title}
  {\bibinfo {title} {\textit{Evolution of Nagaoka phase with kinetic energy
  frustrating hopping}},\ }\href {https://doi.org/10.1103/PhysRevB.95.195103}{\bibfield  {journal} {\bibinfo
  {journal} {Physical Review B}\ }\textbf {\bibinfo {volume} {95}},\ \bibinfo
  {pages} {195103} (\bibinfo {year} {2017})}\BibitemShut {NoStop}%
\bibitem [{\citenamefont {Pereira}\ and\ \citenamefont
  {Mueller}(2025)}]{Pereira_arXiv_2025}%
  \BibitemOpen
  \bibfield  {author} {\bibinfo {author} {\bibfnamefont {D.}~\bibnamefont
  {Pereira}}\ and\ \bibinfo {author} {\bibfnamefont {E.~J.}\ \bibnamefont
  {Mueller}},\ }\bibfield {title}
  {\bibinfo {title} {Kinetic magnetism in the crossover between the square and
  triangular lattice Fermi-Hubbard models},\ }
  \href {https://doi.org/10.1103/v3fr-gch1}
  {\bibfield {journal} {\bibinfo {journal} {Phys. Rev. B}\ }\textbf
  {\bibinfo {volume} {112}},\ \bibinfo {pages} {245120} (\bibinfo {year}
  {2025})}\BibitemShut {NoStop}%
\bibitem [{\citenamefont {Sharma}\ \emph {et~al.}(2026)\citenamefont {Sharma},
  \citenamefont {Peng}, \citenamefont {Sheng}, \citenamefont {Changlani},\ and\
  \citenamefont {Wang}}]{Sharma_arxiv_2025}%
  \BibitemOpen
  \bibfield  {author} {\bibinfo {author} {\bibfnamefont {P.}~\bibnamefont
  {Sharma}}, \bibinfo {author} {\bibfnamefont {Y.}~\bibnamefont {Peng}},
  \bibinfo {author} {\bibfnamefont {D.~N.}\ \bibnamefont {Sheng}}, \bibinfo
  {author} {\bibfnamefont {H.~J.}\ \bibnamefont {Changlani}},\ and\ \bibinfo
  {author} {\bibfnamefont {Y.}~\bibnamefont {Wang}},\ }\bibfield  {title}
  {\bibinfo {title} {Instability of nagaoka state and quantum phase transition
  via kinetic frustration control},\ }\href {https://doi.org/10.1103/z4gs-c6h6}
  {\bibfield  {journal} {\bibinfo  {journal} {Phys. Rev. Res.}\ }\textbf
  {\bibinfo {volume} {8}},\ \bibinfo {pages} {023070} (\bibinfo {year}
  {2026})}\BibitemShut {NoStop}%
\bibitem [{\citenamefont {Zhang}\ \emph {et~al.}(2018)\citenamefont {Zhang},
  \citenamefont {Zhu},\ and\ \citenamefont {Batista}}]{zhang2018pairing}%
  \BibitemOpen
  \bibfield  {author} {\bibinfo {author} {\bibfnamefont {S.-S.}\ \bibnamefont
  {Zhang}}, \bibinfo {author} {\bibfnamefont {W.}~\bibnamefont {Zhu}},\ and\
  \bibinfo {author} {\bibfnamefont {C.~D.}\ \bibnamefont {Batista}},\
  }\bibfield  {title} {\bibinfo {title} {\textit{Pairing from strong repulsion
  in triangular lattice Hubbard model}},\ }\href@noop {} {\bibfield  {journal}
  {\bibinfo  {journal} {Physical Review B}\ }\textbf {\bibinfo {volume} {97}},\
  \bibinfo {pages} {140507} (\bibinfo {year} {2018})}\BibitemShut {NoStop}%
\bibitem [{\citenamefont {Glittum}\ \emph {et~al.}(2025)\citenamefont
  {Glittum}, \citenamefont {{\v{S}}trkalj}, \citenamefont {Prabhakaran},
  \citenamefont {Goddard}, \citenamefont {Batista},\ and\ \citenamefont
  {Castelnovo}}]{glittum2025resonant}%
  \BibitemOpen
  \bibfield  {author} {\bibinfo {author} {\bibfnamefont {C.}~\bibnamefont
  {Glittum}}, \bibinfo {author} {\bibfnamefont {A.}~\bibnamefont
  {{\v{S}}trkalj}}, \bibinfo {author} {\bibfnamefont {D.}~\bibnamefont
  {Prabhakaran}}, \bibinfo {author} {\bibfnamefont {P.~A.}\ \bibnamefont
  {Goddard}}, \bibinfo {author} {\bibfnamefont {C.~D.}\ \bibnamefont
  {Batista}},\ and\ \bibinfo {author} {\bibfnamefont {C.}~\bibnamefont
  {Castelnovo}},\ }\bibfield  {title} {\bibinfo {title} {\textit{A resonant
  valence bond spin liquid in the dilute limit of doped frustrated Mott
  insulators}},\ }\href
  {https://doi.org/https://doi.org/10.1038/s41567-025-02923-8} {\bibfield
  {journal} {\bibinfo  {journal} {Nature Physics}\ ,\ \bibinfo {pages} {1}}
  (\bibinfo {year} {2025})}\BibitemShut {NoStop}%
\bibitem [{\citenamefont {Chen}\ \emph {et~al.}(2025)\citenamefont {Chen},
  \citenamefont {Chen},\ and\ \citenamefont {Zhu}}]{ferro_dmrg_triangular}%
  \BibitemOpen
  \bibfield  {author} {\bibinfo {author} {\bibfnamefont {Q.}~\bibnamefont
  {Chen}}, \bibinfo {author} {\bibfnamefont {S.}~\bibnamefont {Chen}},\ and\
  \bibinfo {author} {\bibfnamefont {Z.}~\bibnamefont {Zhu}},\ }\bibfield
  {title} {\bibinfo {title} {\textit{Geometric frustration assisted kinetic
  ferromagnetism in doped Mott insulators}},\ }\bibfield  {journal} {\bibinfo
  {journal} {Commun Phys}\ }\textbf {\bibinfo {volume} {8}},\ \href
  {https://doi.org/https://doi.org/10.1038/s42005-025-02311-x}
  {https://doi.org/10.1038/s42005-025-02311-x} (\bibinfo {year}
  {2025})\BibitemShut {NoStop}%
\bibitem [{\citenamefont {Prichard}\ \emph {et~al.}(2024)\citenamefont
  {Prichard}, \citenamefont {Spar}, \citenamefont {Morera}, \citenamefont
  {Demler}, \citenamefont {Yan},\ and\ \citenamefont
  {Bakr}}]{prichard2024directly}%
  \BibitemOpen
  \bibfield  {author} {\bibinfo {author} {\bibfnamefont {M.~L.}\ \bibnamefont
  {Prichard}}, \bibinfo {author} {\bibfnamefont {B.~M.}\ \bibnamefont {Spar}},
  \bibinfo {author} {\bibfnamefont {I.}~\bibnamefont {Morera}}, \bibinfo
  {author} {\bibfnamefont {E.}~\bibnamefont {Demler}}, \bibinfo {author}
  {\bibfnamefont {Z.~Z.}\ \bibnamefont {Yan}},\ and\ \bibinfo {author}
  {\bibfnamefont {W.~S.}\ \bibnamefont {Bakr}},\ }\bibfield  {title} {\bibinfo
  {title} {\textit{Directly imaging spin polarons in a kinetically frustrated
  Hubbard system}},\ }\href
  {https://doi.org/https://doi.org/10.1038/s41586-024-07356-6} {\bibfield
  {journal} {\bibinfo  {journal} {Nature}\ }\textbf {\bibinfo {volume} {629}},\
  \bibinfo {pages} {323} (\bibinfo {year} {2024})}\BibitemShut {NoStop}%
\bibitem [{\citenamefont {Lebrat}\ \emph {et~al.}(2024)\citenamefont {Lebrat},
  \citenamefont {Xu}, \citenamefont {Kendrick}, \citenamefont {Kale},
  \citenamefont {Gang}, \citenamefont {Seetharaman}, \citenamefont {Morera},
  \citenamefont {Khatami}, \citenamefont {Demler},\ and\ \citenamefont
  {Greiner}}]{lebrat2024observation}%
  \BibitemOpen
  \bibfield  {author} {\bibinfo {author} {\bibfnamefont {M.}~\bibnamefont
  {Lebrat}}, \bibinfo {author} {\bibfnamefont {M.}~\bibnamefont {Xu}}, \bibinfo
  {author} {\bibfnamefont {L.~H.}\ \bibnamefont {Kendrick}}, \bibinfo {author}
  {\bibfnamefont {A.}~\bibnamefont {Kale}}, \bibinfo {author} {\bibfnamefont
  {Y.}~\bibnamefont {Gang}}, \bibinfo {author} {\bibfnamefont {P.}~\bibnamefont
  {Seetharaman}}, \bibinfo {author} {\bibfnamefont {I.}~\bibnamefont {Morera}},
  \bibinfo {author} {\bibfnamefont {E.}~\bibnamefont {Khatami}}, \bibinfo
  {author} {\bibfnamefont {E.}~\bibnamefont {Demler}},\ and\ \bibinfo {author}
  {\bibfnamefont {M.}~\bibnamefont {Greiner}},\ }\bibfield  {title} {\bibinfo
  {title} {\textit{Observation of Nagaoka polarons in a Fermi--Hubbard quantum
  simulator}},\ }\href
  {https://doi.org/https://doi.org/10.1038/s41586-024-07272-9} {\bibfield
  {journal} {\bibinfo  {journal} {Nature}\ }\textbf {\bibinfo {volume} {629}},\
  \bibinfo {pages} {317} (\bibinfo {year} {2024})}\BibitemShut {NoStop}%
\bibitem [{\citenamefont {Koepsell}\ \emph {et~al.}(2019)\citenamefont
  {Koepsell}, \citenamefont {Vijayan}, \citenamefont {Sompet}, \citenamefont
  {Grusdt}, \citenamefont {Hilker}, \citenamefont {Demler}, \citenamefont
  {Salomon}, \citenamefont {Bloch},\ and\ \citenamefont
  {Gross}}]{koepsell2019imaging}%
  \BibitemOpen
  \bibfield  {author} {\bibinfo {author} {\bibfnamefont {J.}~\bibnamefont
  {Koepsell}}, \bibinfo {author} {\bibfnamefont {J.}~\bibnamefont {Vijayan}},
  \bibinfo {author} {\bibfnamefont {P.}~\bibnamefont {Sompet}}, \bibinfo
  {author} {\bibfnamefont {F.}~\bibnamefont {Grusdt}}, \bibinfo {author}
  {\bibfnamefont {T.~A.}\ \bibnamefont {Hilker}}, \bibinfo {author}
  {\bibfnamefont {E.}~\bibnamefont {Demler}}, \bibinfo {author} {\bibfnamefont
  {G.}~\bibnamefont {Salomon}}, \bibinfo {author} {\bibfnamefont
  {I.}~\bibnamefont {Bloch}},\ and\ \bibinfo {author} {\bibfnamefont
  {C.}~\bibnamefont {Gross}},\ }\bibfield  {title} {\bibinfo {title}
  {\textit{Imaging magnetic polarons in the doped Fermi--Hubbard model}},\
  }\href {https://doi.org/https://doi.org/10.1038/s41586-019-1463-1} {\bibfield
   {journal} {\bibinfo  {journal} {Nature}\ }\textbf {\bibinfo {volume}
  {572}},\ \bibinfo {pages} {358} (\bibinfo {year} {2019})}\BibitemShut
  {NoStop}%
\bibitem [{\citenamefont {White}(1992)}]{White_1992}%
  \BibitemOpen
  \bibfield  {author} {\bibinfo {author} {\bibfnamefont {S.~R.}\ \bibnamefont
  {White}},\ }\bibfield  {title} {\bibinfo {title} {\textit{Density matrix
  formulation for quantum renormalization groups}},\ }\href
  {https://doi.org/10.1103/PhysRevLett.69.2863} {\bibfield  {journal} {\bibinfo
   {journal} {Phys. Rev. Lett.}\ }\textbf {\bibinfo {volume} {69}},\ \bibinfo
  {pages} {2863} (\bibinfo {year} {1992})}\BibitemShut {NoStop}%
\bibitem [{\citenamefont {Stoudenmire}\ and\ \citenamefont
  {White}(2012)}]{Stoudenmire_2012}%
  \BibitemOpen
  \bibfield  {author} {\bibinfo {author} {\bibfnamefont {E.}~\bibnamefont
  {Stoudenmire}}\ and\ \bibinfo {author} {\bibfnamefont {S.~R.}\ \bibnamefont
  {White}},\ }\bibfield  {title} {\bibinfo {title} {\textit{Studying
  Two-Dimensional Systems with the Density Matrix Renormalization Group}},\
  }\href {https://doi.org/10.1146/annurev-conmatphys-020911-125018} {\bibfield
  {journal} {\bibinfo  {journal} {Annual Review of Condensed Matter Physics}\
  }\textbf {\bibinfo {volume} {3}},\ \bibinfo {pages} {111–128} (\bibinfo
  {year} {2012})}\BibitemShut {NoStop}%
\bibitem [{\citenamefont {Li}\ \emph {et~al.}(2014)\citenamefont {Li},
  \citenamefont {Antipov}, \citenamefont {Rubtsov}, \citenamefont {Kirchner},\
  and\ \citenamefont {Hanke}}]{Li2014entropy}%
  \BibitemOpen
  \bibfield  {author} {\bibinfo {author} {\bibfnamefont {G.}~\bibnamefont
  {Li}}, \bibinfo {author} {\bibfnamefont {A.~E.}\ \bibnamefont {Antipov}},
  \bibinfo {author} {\bibfnamefont {A.~N.}\ \bibnamefont {Rubtsov}}, \bibinfo
  {author} {\bibfnamefont {S.}~\bibnamefont {Kirchner}},\ and\ \bibinfo
  {author} {\bibfnamefont {W.}~\bibnamefont {Hanke}},\ }\bibfield  {title}
  {\bibinfo {title} {\textit{Competing phases of the Hubbard model on a
  triangular lattice: Insights from the entropy}},\ }\href
  {https://doi.org/10.1103/PhysRevB.89.161118} {\bibfield  {journal} {\bibinfo
  {journal} {Phys. Rev. B}\ }\textbf {\bibinfo {volume} {89}},\ \bibinfo
  {pages} {161118} (\bibinfo {year} {2014})}\BibitemShut {NoStop}%
\bibitem [{\citenamefont {Merino}\ \emph {et~al.}(2006)\citenamefont {Merino},
  \citenamefont {Powell},\ and\ \citenamefont
  {McKenzie}}]{merino2006ferromagnetism}%
  \BibitemOpen
  \bibfield  {author} {\bibinfo {author} {\bibfnamefont {J.}~\bibnamefont
  {Merino}}, \bibinfo {author} {\bibfnamefont {B.}~\bibnamefont {Powell}},\
  and\ \bibinfo {author} {\bibfnamefont {R.~H.}\ \bibnamefont {McKenzie}},\
  }\bibfield  {title} {\bibinfo {title} {\textit{Ferromagnetism, paramagnetism,
  and a Curie-Weiss metal in an electron-doped Hubbard model on a triangular
  lattice}},\ }\href {https://doi.org/10.1103/PhysRevB.73.235107} {\bibfield
  {journal} {\bibinfo  {journal} {Physical Review B—Condensed Matter and
  Materials Physics}\ }\textbf {\bibinfo {volume} {73}},\ \bibinfo {pages}
  {235107} (\bibinfo {year} {2006})}\BibitemShut {NoStop}%
\bibitem [{\citenamefont {Yoshioka}\ \emph {et~al.}(2009)\citenamefont
  {Yoshioka}, \citenamefont {Koga},\ and\ \citenamefont
  {Kawakami}}]{yoshioka2009quantum}%
  \BibitemOpen
  \bibfield  {author} {\bibinfo {author} {\bibfnamefont {T.}~\bibnamefont
  {Yoshioka}}, \bibinfo {author} {\bibfnamefont {A.}~\bibnamefont {Koga}},\
  and\ \bibinfo {author} {\bibfnamefont {N.}~\bibnamefont {Kawakami}},\
  }\bibfield  {title} {\bibinfo {title} {\textit{Quantum phase transitions in
  the Hubbard model on a triangular lattice}},\ }\href
  {https://doi.org/10.1103/PhysRevLett.103.036401} {\bibfield  {journal}
  {\bibinfo  {journal} {Physical review letters}\ }\textbf {\bibinfo {volume}
  {103}},\ \bibinfo {pages} {036401} (\bibinfo {year} {2009})}\BibitemShut
  {NoStop}%
\bibitem [{\citenamefont {Sahebsara}\ and\ \citenamefont
  {S{\'e}n{\'e}chal}(2008)}]{sahebsara2008hubbard}%
  \BibitemOpen
  \bibfield  {author} {\bibinfo {author} {\bibfnamefont {P.}~\bibnamefont
  {Sahebsara}}\ and\ \bibinfo {author} {\bibfnamefont {D.}~\bibnamefont
  {S{\'e}n{\'e}chal}},\ }\bibfield  {title} {\bibinfo {title} {\textit{Hubbard
  model on the triangular lattice: Spiral order and spin liquid}},\ }\href
  {https://doi.org/10.1103/PhysRevLett.100.136402} {\bibfield  {journal}
  {\bibinfo  {journal} {Physical review letters}\ }\textbf {\bibinfo {volume}
  {100}},\ \bibinfo {pages} {136402} (\bibinfo {year} {2008})}\BibitemShut
  {NoStop}%
\bibitem [{\citenamefont {Wietek}\ \emph {et~al.}(2021)\citenamefont {Wietek},
  \citenamefont {Rossi}, \citenamefont {{\v{S}}imkovic~IV}, \citenamefont
  {Klett}, \citenamefont {Hansmann}, \citenamefont {Ferrero}, \citenamefont
  {Stoudenmire}, \citenamefont {Sch{\"a}fer},\ and\ \citenamefont
  {Georges}}]{wietek2021mott}%
  \BibitemOpen
  \bibfield  {author} {\bibinfo {author} {\bibfnamefont {A.}~\bibnamefont
  {Wietek}}, \bibinfo {author} {\bibfnamefont {R.}~\bibnamefont {Rossi}},
  \bibinfo {author} {\bibfnamefont {F.}~\bibnamefont {{\v{S}}imkovic~IV}},
  \bibinfo {author} {\bibfnamefont {M.}~\bibnamefont {Klett}}, \bibinfo
  {author} {\bibfnamefont {P.}~\bibnamefont {Hansmann}}, \bibinfo {author}
  {\bibfnamefont {M.}~\bibnamefont {Ferrero}}, \bibinfo {author} {\bibfnamefont
  {E.~M.}\ \bibnamefont {Stoudenmire}}, \bibinfo {author} {\bibfnamefont
  {T.}~\bibnamefont {Sch{\"a}fer}},\ and\ \bibinfo {author} {\bibfnamefont
  {A.}~\bibnamefont {Georges}},\ }\bibfield  {title} {\bibinfo {title}
  {\textit{Mott insulating states with competing orders in the triangular
  lattice Hubbard model}},\ }\href {https://doi.org/10.1103/PhysRevX.11.041013}
  {\bibfield  {journal} {\bibinfo  {journal} {Physical Review X}\ }\textbf
  {\bibinfo {volume} {11}},\ \bibinfo {pages} {041013} (\bibinfo {year}
  {2021})}\BibitemShut {NoStop}%
\bibitem [{\citenamefont {Shirakawa}\ \emph {et~al.}(2017)\citenamefont
  {Shirakawa}, \citenamefont {Tohyama}, \citenamefont {Kokalj}, \citenamefont
  {Sota},\ and\ \citenamefont {Yunoki}}]{beyond_hesinberg2017}%
  \BibitemOpen
  \bibfield  {author} {\bibinfo {author} {\bibfnamefont {T.}~\bibnamefont
  {Shirakawa}}, \bibinfo {author} {\bibfnamefont {T.}~\bibnamefont {Tohyama}},
  \bibinfo {author} {\bibfnamefont {J.}~\bibnamefont {Kokalj}}, \bibinfo
  {author} {\bibfnamefont {S.}~\bibnamefont {Sota}},\ and\ \bibinfo {author}
  {\bibfnamefont {S.}~\bibnamefont {Yunoki}},\ }\bibfield  {title} {\bibinfo
  {title} {\textit{Ground-state phase diagram of the triangular lattice Hubbard
  model by the density-matrix renormalization group method}},\ }\href
  {https://doi.org/10.1103/PhysRevB.96.205130} {\bibfield  {journal} {\bibinfo
  {journal} {Phys. Rev. B}\ }\textbf {\bibinfo {volume} {96}},\ \bibinfo
  {pages} {205130} (\bibinfo {year} {2017})}\BibitemShut {NoStop}%
\bibitem [{\citenamefont {Lee}\ \emph {et~al.}(2023)\citenamefont {Lee},
  \citenamefont {Sharma}, \citenamefont {Vafek},\ and\ \citenamefont
  {Changlani}}]{lee2023triangular}%
  \BibitemOpen
  \bibfield  {author} {\bibinfo {author} {\bibfnamefont {K.}~\bibnamefont
  {Lee}}, \bibinfo {author} {\bibfnamefont {P.}~\bibnamefont {Sharma}},
  \bibinfo {author} {\bibfnamefont {O.}~\bibnamefont {Vafek}},\ and\ \bibinfo
  {author} {\bibfnamefont {H.~J.}\ \bibnamefont {Changlani}},\ }\bibfield
  {title} {\bibinfo {title} {\textit{Triangular lattice Hubbard model physics
  at intermediate temperatures}},\ }\href
  {https://doi.org/10.1103/PhysRevB.107.235105} {\bibfield  {journal} {\bibinfo
   {journal} {Physical Review B}\ }\textbf {\bibinfo {volume} {107}},\ \bibinfo
  {pages} {235105} (\bibinfo {year} {2023})}\BibitemShut {NoStop}%
\bibitem [{\citenamefont {Morera}\ \emph {et~al.}(2023)\citenamefont {Morera},
  \citenamefont {Kan\'asz-Nagy}, \citenamefont {Smolenski}, \citenamefont
  {Ciorciaro}, \citenamefont {Imamo\ifmmode~\breve{g}\else \u{g}\fi{}lu},\ and\
  \citenamefont {Demler}}]{Morera_2023}%
  \BibitemOpen
  \bibfield  {author} {\bibinfo {author} {\bibfnamefont {I.}~\bibnamefont
  {Morera}}, \bibinfo {author} {\bibfnamefont {M.}~\bibnamefont
  {Kan\'asz-Nagy}}, \bibinfo {author} {\bibfnamefont {T.}~\bibnamefont
  {Smolenski}}, \bibinfo {author} {\bibfnamefont {L.}~\bibnamefont
  {Ciorciaro}}, \bibinfo {author} {\bibfnamefont {A.~m.~c.}\ \bibnamefont
  {Imamo\ifmmode~\breve{g}\else \u{g}\fi{}lu}},\ and\ \bibinfo {author}
  {\bibfnamefont {E.}~\bibnamefont {Demler}},\ }\bibfield  {title} {\bibinfo
  {title} {\textit{High-temperature kinetic magnetism in triangular
  lattices}},\ }\href {https://doi.org/10.1103/PhysRevResearch.5.L022048}
  {\bibfield  {journal} {\bibinfo  {journal} {Phys. Rev. Res.}\ }\textbf
  {\bibinfo {volume} {5}},\ \bibinfo {pages} {L022048} (\bibinfo {year}
  {2023})}\BibitemShut {NoStop}%
\bibitem [{\citenamefont {Xu}\ \emph {et~al.}(2023)\citenamefont {Xu},
  \citenamefont {Kendrick}, \citenamefont {Kale}, \citenamefont {Gang},
  \citenamefont {Ji}, \citenamefont {Scalettar}, \citenamefont {Lebrat},\ and\
  \citenamefont {Greiner}}]{greinerXu}%
  \BibitemOpen
  \bibfield  {author} {\bibinfo {author} {\bibfnamefont {M.}~\bibnamefont
  {Xu}}, \bibinfo {author} {\bibfnamefont {L.~H.}\ \bibnamefont {Kendrick}},
  \bibinfo {author} {\bibfnamefont {A.}~\bibnamefont {Kale}}, \bibinfo {author}
  {\bibfnamefont {Y.}~\bibnamefont {Gang}}, \bibinfo {author} {\bibfnamefont
  {G.}~\bibnamefont {Ji}}, \bibinfo {author} {\bibfnamefont {R.~T.}\
  \bibnamefont {Scalettar}}, \bibinfo {author} {\bibfnamefont {M.}~\bibnamefont
  {Lebrat}},\ and\ \bibinfo {author} {\bibfnamefont {M.}~\bibnamefont
  {Greiner}},\ }\bibfield  {title} {\bibinfo {title} {\textit{Frustration- and
  doping-induced magnetism in a Fermi{\textendash}Hubbard simulator}},\
  }\bibfield  {journal} {\bibinfo  {journal} {Nature}\ }\textbf {\bibinfo
  {volume} {620}},\ \href {https://doi.org/10.1038/s41586-023-06280-5}
  {10.1038/s41586-023-06280-5} (\bibinfo {year} {(2023)})\BibitemShut {NoStop}%
\bibitem [{\citenamefont {Mazurenko}\ \emph {et~al.}(2017)\citenamefont
  {Mazurenko}, \citenamefont {Chiu}, \citenamefont {Ji}, \citenamefont
  {Parsons}, \citenamefont {Kan{\'a}sz-Nagy}, \citenamefont {Schmidt},
  \citenamefont {Grusdt}, \citenamefont {Demler}, \citenamefont {Greif},\ and\
  \citenamefont {Greiner}}]{mazurenko2017cold}%
  \BibitemOpen
  \bibfield  {author} {\bibinfo {author} {\bibfnamefont {A.}~\bibnamefont
  {Mazurenko}}, \bibinfo {author} {\bibfnamefont {C.~S.}\ \bibnamefont {Chiu}},
  \bibinfo {author} {\bibfnamefont {G.}~\bibnamefont {Ji}}, \bibinfo {author}
  {\bibfnamefont {M.~F.}\ \bibnamefont {Parsons}}, \bibinfo {author}
  {\bibfnamefont {M.}~\bibnamefont {Kan{\'a}sz-Nagy}}, \bibinfo {author}
  {\bibfnamefont {R.}~\bibnamefont {Schmidt}}, \bibinfo {author} {\bibfnamefont
  {F.}~\bibnamefont {Grusdt}}, \bibinfo {author} {\bibfnamefont
  {E.}~\bibnamefont {Demler}}, \bibinfo {author} {\bibfnamefont
  {D.}~\bibnamefont {Greif}},\ and\ \bibinfo {author} {\bibfnamefont
  {M.}~\bibnamefont {Greiner}},\ }\bibfield  {title} {\bibinfo {title}
  {\textit{A cold-atom Fermi--Hubbard antiferromagnet}},\ }\href
  {https://doi.org/https://doi.org/10.1038/nature22362} {\bibfield  {journal}
  {\bibinfo  {journal} {Nature}\ }\textbf {\bibinfo {volume} {545}},\ \bibinfo
  {pages} {462} (\bibinfo {year} {2017})}\BibitemShut {NoStop}%
\bibitem [{\citenamefont {Qiao}\ \emph {et~al.}(2025)\citenamefont {Qiao},
  \citenamefont {Martin}, \citenamefont {Homeier}, \citenamefont {Morera},
  \citenamefont {Gély}, \citenamefont {Klein}, \citenamefont {Chew},
  \citenamefont {Barredo}, \citenamefont {Lahaye}, \citenamefont {Demler},\
  and\ \citenamefont
  {Browaeys}}]{qiao2025kineticallyinducedboundstatesfrustrated}%
  \BibitemOpen
  \bibfield  {author} {\bibinfo {author} {\bibfnamefont {M.}~\bibnamefont
  {Qiao}}, \bibinfo {author} {\bibfnamefont {R.}~\bibnamefont {Martin}},
  \bibinfo {author} {\bibfnamefont {L.}~\bibnamefont {Homeier}}, \bibinfo
  {author} {\bibfnamefont {I.}~\bibnamefont {Morera}}, \bibinfo {author}
  {\bibfnamefont {B.}~\bibnamefont {Gély}}, \bibinfo {author} {\bibfnamefont
  {L.}~\bibnamefont {Klein}}, \bibinfo {author} {\bibfnamefont {Y.~T.}\
  \bibnamefont {Chew}}, \bibinfo {author} {\bibfnamefont {D.}~\bibnamefont
  {Barredo}}, \bibinfo {author} {\bibfnamefont {T.}~\bibnamefont {Lahaye}},
  \bibinfo {author} {\bibfnamefont {E.}~\bibnamefont {Demler}},\ and\ \bibinfo
  {author} {\bibfnamefont {A.}~\bibnamefont {Browaeys}},\ }\href
  {https://doi.org/https://doi.org/10.48550/arXiv.2510.17183} {\bibinfo {title}
  {\textit{Kinetically-induced bound states in a frustrated Rydberg tweezer
  array}}} (\bibinfo {year} {2025}),\ \Eprint
  {https://arxiv.org/abs/2510.17183} {arXiv:2510.17183 [quant-ph]} \BibitemShut
  {NoStop}%
\bibitem [{\citenamefont {Tang}\ \emph {et~al.}(2022)\citenamefont {Tang},
  \citenamefont {Li}, \citenamefont {Li},\ and\ \citenamefont {et~al.}}]{tang}%
  \BibitemOpen
  \bibfield  {author} {\bibinfo {author} {\bibfnamefont {Y.}~\bibnamefont
  {Tang}}, \bibinfo {author} {\bibfnamefont {L.}~\bibnamefont {Li}}, \bibinfo
  {author} {\bibfnamefont {T.}~\bibnamefont {Li}},\ and\ \bibinfo {author}
  {\bibnamefont {et~al.}},\ }\bibfield  {title} {\bibinfo {title}
  {\textit{Simulation of Hubbard model physics in WSe2/WS2 moiré
  superlattices}},\ }\bibfield  {journal} {\bibinfo  {journal} {Nature}\
  }\textbf {\bibinfo {volume} {579}},\ \href
  {https://doi.org/https://doi.org/10.1038/s41586-020-2085-3}
  {https://doi.org/10.1038/s41586-020-2085-3} (\bibinfo {year}
  {(2022)})\BibitemShut {NoStop}%
\bibitem [{\citenamefont {Ciorciaro}\ \emph {et~al.}(2023)\citenamefont
  {Ciorciaro}, \citenamefont {Smole{\'n}ski}, \citenamefont {Morera},
  \citenamefont {Kiper}, \citenamefont {Hiestand}, \citenamefont {Kroner},
  \citenamefont {Zhang}, \citenamefont {Watanabe}, \citenamefont {Taniguchi},
  \citenamefont {Demler} \emph {et~al.}}]{ciorciaro2023kinetic}%
  \BibitemOpen
  \bibfield  {author} {\bibinfo {author} {\bibfnamefont {L.}~\bibnamefont
  {Ciorciaro}}, \bibinfo {author} {\bibfnamefont {T.}~\bibnamefont
  {Smole{\'n}ski}}, \bibinfo {author} {\bibfnamefont {I.}~\bibnamefont
  {Morera}}, \bibinfo {author} {\bibfnamefont {N.}~\bibnamefont {Kiper}},
  \bibinfo {author} {\bibfnamefont {S.}~\bibnamefont {Hiestand}}, \bibinfo
  {author} {\bibfnamefont {M.}~\bibnamefont {Kroner}}, \bibinfo {author}
  {\bibfnamefont {Y.}~\bibnamefont {Zhang}}, \bibinfo {author} {\bibfnamefont
  {K.}~\bibnamefont {Watanabe}}, \bibinfo {author} {\bibfnamefont
  {T.}~\bibnamefont {Taniguchi}}, \bibinfo {author} {\bibfnamefont
  {E.}~\bibnamefont {Demler}}, \emph {et~al.},\ }\bibfield  {title} {\bibinfo
  {title} {\textit{Kinetic magnetism in triangular moir{\'e} materials}},\
  }\href {https://doi.org/https://doi.org/10.1038/s41586-023-06633-0}
  {\bibfield  {journal} {\bibinfo  {journal} {Nature}\ }\textbf {\bibinfo
  {volume} {623}},\ \bibinfo {pages} {509} (\bibinfo {year}
  {2023})}\BibitemShut {NoStop}%
\bibitem [{\citenamefont {Tao}\ \emph {et~al.}(2024)\citenamefont {Tao},
  \citenamefont {Zhao}, \citenamefont {Shen}, \citenamefont {Li}, \citenamefont
  {Kn{\"u}ppel}, \citenamefont {Watanabe}, \citenamefont {Taniguchi},
  \citenamefont {Shan},\ and\ \citenamefont {Mak}}]{tao2024observation}%
  \BibitemOpen
  \bibfield  {author} {\bibinfo {author} {\bibfnamefont {Z.}~\bibnamefont
  {Tao}}, \bibinfo {author} {\bibfnamefont {W.}~\bibnamefont {Zhao}}, \bibinfo
  {author} {\bibfnamefont {B.}~\bibnamefont {Shen}}, \bibinfo {author}
  {\bibfnamefont {T.}~\bibnamefont {Li}}, \bibinfo {author} {\bibfnamefont
  {P.}~\bibnamefont {Kn{\"u}ppel}}, \bibinfo {author} {\bibfnamefont
  {K.}~\bibnamefont {Watanabe}}, \bibinfo {author} {\bibfnamefont
  {T.}~\bibnamefont {Taniguchi}}, \bibinfo {author} {\bibfnamefont
  {J.}~\bibnamefont {Shan}},\ and\ \bibinfo {author} {\bibfnamefont {K.~F.}\
  \bibnamefont {Mak}},\ }\bibfield  {title} {\bibinfo {title}
  {\textit{Observation of spin polarons in a frustrated moir{\'e} Hubbard
  system}},\ }\href@noop {} {\bibfield  {journal} {\bibinfo  {journal} {Nature
  Physics}\ }\textbf {\bibinfo {volume} {20}},\ \bibinfo {pages} {783}
  (\bibinfo {year} {2024})}\BibitemShut {NoStop}%
\bibitem [{\citenamefont {Carlstr\"om}(2022)}]{Carlstrom_2022}%
  \BibitemOpen
  \bibfield  {author} {\bibinfo {author} {\bibfnamefont {J.}~\bibnamefont
  {Carlstr\"om}},\ }\bibfield  {title} {\bibinfo {title} {In situ controllable
  magnetic phases in doped twisted bilayer transition metal dichalcogenides},\
  }\href {https://doi.org/10.1103/PhysRevResearch.4.043126} {\bibfield
  {journal} {\bibinfo  {journal} {Phys. Rev. Res.}\ }\textbf {\bibinfo {volume}
  {4}},\ \bibinfo {pages} {043126} (\bibinfo {year} {2022})}\BibitemShut
  {NoStop}%
\bibitem [{\citenamefont {Barford}\ and\ \citenamefont
  {Kim}(1991)}]{barford1991spinless}%
  \BibitemOpen
  \bibfield  {author} {\bibinfo {author} {\bibfnamefont {W.}~\bibnamefont
  {Barford}}\ and\ \bibinfo {author} {\bibfnamefont {J.~H.}\ \bibnamefont
  {Kim}},\ }\bibfield  {title} {\bibinfo {title} {\textit{Spinless fermions on
  frustrated lattices in a magnetic field}},\ }\href
  {https://doi.org/10.1103/PhysRevB.43.559} {\bibfield  {journal} {\bibinfo
  {journal} {Physical Review B}\ }\textbf {\bibinfo {volume} {43}},\ \bibinfo
  {pages} {559} (\bibinfo {year} {1991})}\BibitemShut {NoStop}%
\bibitem [{\citenamefont {Bernu}\ \emph {et~al.}(1992)\citenamefont {Bernu},
  \citenamefont {Lhuillier},\ and\ \citenamefont {Pierre}}]{Bernu_1992}%
  \BibitemOpen
  \bibfield  {author} {\bibinfo {author} {\bibfnamefont {B.}~\bibnamefont
  {Bernu}}, \bibinfo {author} {\bibfnamefont {C.}~\bibnamefont {Lhuillier}},\
  and\ \bibinfo {author} {\bibfnamefont {L.}~\bibnamefont {Pierre}},\
  }\bibfield  {title} {\bibinfo {title} {\textit{Signature of N\'eel order in
  exact spectra of quantum antiferromagnets on finite lattices}},\ }\href
  {https://doi.org/10.1103/PhysRevLett.69.2590} {\bibfield  {journal} {\bibinfo
   {journal} {Phys. Rev. Lett.}\ }\textbf {\bibinfo {volume} {69}},\ \bibinfo
  {pages} {2590} (\bibinfo {year} {1992})}\BibitemShut {NoStop}%
\bibitem [{\citenamefont {Liu}\ \emph {et~al.}(2012)\citenamefont {Liu},
  \citenamefont {Yao}, \citenamefont {Berg}, \citenamefont {White},\ and\
  \citenamefont {Kivelson}}]{squareinfU}%
  \BibitemOpen
  \bibfield  {author} {\bibinfo {author} {\bibfnamefont {L.}~\bibnamefont
  {Liu}}, \bibinfo {author} {\bibfnamefont {H.}~\bibnamefont {Yao}}, \bibinfo
  {author} {\bibfnamefont {E.}~\bibnamefont {Berg}}, \bibinfo {author}
  {\bibfnamefont {S.~R.}\ \bibnamefont {White}},\ and\ \bibinfo {author}
  {\bibfnamefont {S.~A.}\ \bibnamefont {Kivelson}},\ }\bibfield  {title}
  {\bibinfo {title} {\textit{Phases of the infinite U Hubbard model on square
  lattices}},\ }\href {https://doi.org/10.1103/PhysRevLett.108.126406}
  {\bibfield  {journal} {\bibinfo  {journal} {Phys. Rev. Lett.}\ }\textbf
  {\bibinfo {volume} {108}},\ \bibinfo {pages} {126406} (\bibinfo {year}
  {2012})}\BibitemShut {NoStop}%
\bibitem [{\citenamefont {Coleman}(2015)}]{Coleman2015}%
  \BibitemOpen
  \bibfield {author} {\bibinfo {author} {\bibfnamefont {P.}~\bibnamefont
  {Coleman}},\ }
  \href {https://doi.org/10.1017/CBO9781139020916}
  {\bibfield {title} {\bibinfo {title} {\textit{Introduction to Many-Body
  Physics}}}}
  \ (\bibinfo {publisher} {Cambridge University Press},\
  \bibinfo {year} {2015})\BibitemShut {NoStop}%
\bibitem [{\citenamefont {Fratini}\ \emph {et~al.}(2026)\citenamefont
  {Fratini}, \citenamefont {Duchemin}, \citenamefont {Ralko},\ and\
  \citenamefont {Ciuchi}}]{Fratini_2025}%
  \BibitemOpen
  \bibfield {author} {\bibinfo {author} {\bibfnamefont {S.}~\bibnamefont
  {Fratini}}, \bibinfo {author} {\bibfnamefont {I.}~\bibnamefont {Duchemin}},
  \bibinfo {author} {\bibfnamefont {A.}~\bibnamefont {Ralko}},\ and\
  \bibinfo {author} {\bibfnamefont {S.}~\bibnamefont {Ciuchi}},\
  }\href@noop {} {\bibinfo {title} {\textit{Strange metal transport from
  coupling to fluctuating spins}}} (\bibinfo {year} {2026}),\ \Eprint
  {https://arxiv.org/abs/2412.04322} {arXiv:2412.04322 [cond-mat.str-el]}
  \BibitemShut {NoStop}%
\bibitem [{\citenamefont {Feynman}(1954)}]{feynman}%
  \BibitemOpen
  \bibfield  {author} {\bibinfo {author} {\bibfnamefont {R.~P.}\ \bibnamefont
  {Feynman}},\ }\bibfield  {title} {\bibinfo {title} {\textit{Atomic Theory of
  the Two-Fluid Model of Liquid Helium}},\ }\bibfield  {journal} {\bibinfo
  {journal} {Phys. Rev.}\ }\textbf {\bibinfo {volume} {94}},\ \href
  {https://doi.org/10.1103/PhysRev.94.262} {10.1103/PhysRev.94.262} (\bibinfo
  {year} {(1954)})\BibitemShut {NoStop}%
\bibitem [{\citenamefont {Capello}\ \emph {et~al.}(2006)\citenamefont
  {Capello}, \citenamefont {Becca}, \citenamefont {Yunoki},\ and\ \citenamefont
  {Sorella}}]{Sorella}%
  \BibitemOpen
  \bibfield  {author} {\bibinfo {author} {\bibfnamefont {M.}~\bibnamefont
  {Capello}}, \bibinfo {author} {\bibfnamefont {F.}~\bibnamefont {Becca}},
  \bibinfo {author} {\bibfnamefont {S.}~\bibnamefont {Yunoki}},\ and\ \bibinfo
  {author} {\bibfnamefont {S.}~\bibnamefont {Sorella}},\ }\bibfield  {title}
  {\bibinfo {title} {\textit{Unconventional metal-insulator transition in two
  dimensions}},\ }\bibfield  {journal} {\bibinfo  {journal} {Phys. Rev. B}\
  }\textbf {\bibinfo {volume} {73}},\ \href
  {https://doi.org/10.1103/PhysRevB.73.245116} {10.1103/PhysRevB.73.245116}
  (\bibinfo {year} {(2006)})\BibitemShut {NoStop}%
\bibitem [{\citenamefont {Shackleton}\ \emph {et~al.}(2021)\citenamefont
  {Shackleton}, \citenamefont {Wietek}, \citenamefont {Georges},\ and\
  \citenamefont {Sachdev}}]{Shackleton_PRL_2021}%
  \BibitemOpen
  \bibfield  {author} {\bibinfo {author} {\bibfnamefont {H.}~\bibnamefont
  {Shackleton}}, \bibinfo {author} {\bibfnamefont {A.}~\bibnamefont {Wietek}},
  \bibinfo {author} {\bibfnamefont {A.}~\bibnamefont {Georges}},\ and\ \bibinfo
  {author} {\bibfnamefont {S.}~\bibnamefont {Sachdev}},\ }\bibfield  {title}
  {\bibinfo {title} {\textit{Quantum Phase Transition at Nonzero Doping in a
  Random $t\text{\ensuremath{-}}J$ Model}},\ }\href
  {https://doi.org/10.1103/PhysRevLett.126.136602} {\bibfield  {journal}
  {\bibinfo  {journal} {Phys. Rev. Lett.}\ }\textbf {\bibinfo {volume} {126}},\
  \bibinfo {pages} {136602} (\bibinfo {year} {2021})}\BibitemShut {NoStop}%
\bibitem [{\citenamefont {Kumar}\ \emph {et~al.}(2022)\citenamefont {Kumar},
  \citenamefont {Sachdev},\ and\ \citenamefont
  {Tripathi}}]{kumar_quasiparticle_t_J_2022}%
  \BibitemOpen
  \bibfield  {author} {\bibinfo {author} {\bibfnamefont {A.}~\bibnamefont
  {Kumar}}, \bibinfo {author} {\bibfnamefont {S.}~\bibnamefont {Sachdev}},\
  and\ \bibinfo {author} {\bibfnamefont {V.}~\bibnamefont {Tripathi}},\
  }\bibfield  {title} {\bibinfo {title} {\textit{Quasiparticle metamorphosis in
  the random $t\text{\ensuremath{-}}J$ model}},\ }\href
  {https://doi.org/10.1103/PhysRevB.106.L081120} {\bibfield  {journal}
  {\bibinfo  {journal} {Phys. Rev. B}\ }\textbf {\bibinfo {volume} {106}},\
  \bibinfo {pages} {L081120} (\bibinfo {year} {2022})}\BibitemShut {NoStop}%
\bibitem [{\citenamefont {Szasz}\ \emph {et~al.}(2020)\citenamefont {Szasz},
  \citenamefont {Motruk}, \citenamefont {Zaletel},\ and\ \citenamefont
  {Moore}}]{Zaletel_SpinLiquid}%
  \BibitemOpen
  \bibfield  {author} {\bibinfo {author} {\bibfnamefont {A.}~\bibnamefont
  {Szasz}}, \bibinfo {author} {\bibfnamefont {J.}~\bibnamefont {Motruk}},
  \bibinfo {author} {\bibfnamefont {M.~P.}\ \bibnamefont {Zaletel}},\ and\
  \bibinfo {author} {\bibfnamefont {J.~E.}\ \bibnamefont {Moore}},\ }\bibfield
  {title} {\bibinfo {title} {\textit{Chiral Spin Liquid Phase of the Triangular
  Lattice Hubbard Model: A Density Matrix Renormalization Group Study}},\
  }\href {https://doi.org/10.1103/PhysRevX.10.021042} {\bibfield  {journal}
  {\bibinfo  {journal} {Phys. Rev. X}\ }\textbf {\bibinfo {volume} {10}},\
  \bibinfo {pages} {021042} (\bibinfo {year} {2020})}\BibitemShut {NoStop}%
\bibitem [{\citenamefont {Alexandradinata}\ \emph {et~al.}(2025)\citenamefont
  {Alexandradinata}, \citenamefont {Armitage}, \citenamefont {Baydin},
  \citenamefont {Bi}, \citenamefont {Cao}, \citenamefont {Changlani},
  \citenamefont {Chertkov}, \citenamefont {da~Silva~Neto}, \citenamefont
  {Delacretaz}, \citenamefont {Baggari}, \citenamefont {Ferguson},
  \citenamefont {Gannon}, \citenamefont {Ghorashi}, \citenamefont {Goodge},
  \citenamefont {Goulko}, \citenamefont {Grissonnanche}, \citenamefont
  {Hallas}, \citenamefont {Hayes}, \citenamefont {He}, \citenamefont {Huang},
  \citenamefont {Kogar}, \citenamefont {Kumah}, \citenamefont {Lee},
  \citenamefont {Legros}, \citenamefont {Mahmood}, \citenamefont {Maximenko},
  \citenamefont {Pellatz}, \citenamefont {Polshyn}, \citenamefont {Sarkar},
  \citenamefont {Scheie}, \citenamefont {Seyler}, \citenamefont {Shi},
  \citenamefont {Skinner}, \citenamefont {Steinke}, \citenamefont
  {Thirunavukkuarasu}, \citenamefont {Trevisan}, \citenamefont {Vogl},
  \citenamefont {Volkov}, \citenamefont {Wang}, \citenamefont {Wang},
  \citenamefont {Wei}, \citenamefont {Wei}, \citenamefont {Yang}, \citenamefont
  {Zhang}, \citenamefont {Zhang}, \citenamefont {Zhao},\ and\ \citenamefont
  {Zong}}]{Future_SC_2025}%
  \BibitemOpen
  \bibfield  {author} {\bibinfo {author} {\bibfnamefont {A.}~\bibnamefont
  {Alexandradinata}}, \bibinfo {author} {\bibfnamefont {N.~P.}\ \bibnamefont
  {Armitage}}, \bibinfo {author} {\bibfnamefont {A.}~\bibnamefont {Baydin}},
  \bibinfo {author} {\bibfnamefont {W.}~\bibnamefont {Bi}}, \bibinfo {author}
  {\bibfnamefont {Y.}~\bibnamefont {Cao}}, \bibinfo {author} {\bibfnamefont
  {H.~J.}\ \bibnamefont {Changlani}}, \bibinfo {author} {\bibfnamefont
  {E.}~\bibnamefont {Chertkov}}, \bibinfo {author} {\bibfnamefont {E.~H.}\
  \bibnamefont {da~Silva~Neto}}, \bibinfo {author} {\bibfnamefont
  {L.}~\bibnamefont {Delacretaz}}, \bibinfo {author} {\bibfnamefont {I.~E.}\
  \bibnamefont {Baggari}}, \bibinfo {author} {\bibfnamefont {G.~M.}\
  \bibnamefont {Ferguson}}, \bibinfo {author} {\bibfnamefont {W.~J.}\
  \bibnamefont {Gannon}}, \bibinfo {author} {\bibfnamefont {S.~A.~A.}\
  \bibnamefont {Ghorashi}}, \bibinfo {author} {\bibfnamefont {B.~H.}\
  \bibnamefont {Goodge}}, \bibinfo {author} {\bibfnamefont {O.}~\bibnamefont
  {Goulko}}, \bibinfo {author} {\bibfnamefont {G.}~\bibnamefont
  {Grissonnanche}}, \bibinfo {author} {\bibfnamefont {A.}~\bibnamefont
  {Hallas}}, \bibinfo {author} {\bibfnamefont {I.~M.}\ \bibnamefont {Hayes}},
  \bibinfo {author} {\bibfnamefont {Y.}~\bibnamefont {He}}, \bibinfo {author}
  {\bibfnamefont {E.~W.}\ \bibnamefont {Huang}}, \bibinfo {author}
  {\bibfnamefont {A.}~\bibnamefont {Kogar}}, \bibinfo {author} {\bibfnamefont
  {D.}~\bibnamefont {Kumah}}, \bibinfo {author} {\bibfnamefont {J.~Y.}\
  \bibnamefont {Lee}}, \bibinfo {author} {\bibfnamefont {A.}~\bibnamefont
  {Legros}}, \bibinfo {author} {\bibfnamefont {F.}~\bibnamefont {Mahmood}},
  \bibinfo {author} {\bibfnamefont {Y.}~\bibnamefont {Maximenko}}, \bibinfo
  {author} {\bibfnamefont {N.}~\bibnamefont {Pellatz}}, \bibinfo {author}
  {\bibfnamefont {H.}~\bibnamefont {Polshyn}}, \bibinfo {author} {\bibfnamefont
  {T.}~\bibnamefont {Sarkar}}, \bibinfo {author} {\bibfnamefont
  {A.}~\bibnamefont {Scheie}}, \bibinfo {author} {\bibfnamefont {K.~L.}\
  \bibnamefont {Seyler}}, \bibinfo {author} {\bibfnamefont {Z.}~\bibnamefont
  {Shi}}, \bibinfo {author} {\bibfnamefont {B.}~\bibnamefont {Skinner}},
  \bibinfo {author} {\bibfnamefont {L.}~\bibnamefont {Steinke}}, \bibinfo
  {author} {\bibfnamefont {K.}~\bibnamefont {Thirunavukkuarasu}}, \bibinfo
  {author} {\bibfnamefont {T.~V.}\ \bibnamefont {Trevisan}}, \bibinfo {author}
  {\bibfnamefont {M.}~\bibnamefont {Vogl}}, \bibinfo {author} {\bibfnamefont
  {P.~A.}\ \bibnamefont {Volkov}}, \bibinfo {author} {\bibfnamefont
  {Y.}~\bibnamefont {Wang}}, \bibinfo {author} {\bibfnamefont {Y.}~\bibnamefont
  {Wang}}, \bibinfo {author} {\bibfnamefont {D.}~\bibnamefont {Wei}}, \bibinfo
  {author} {\bibfnamefont {K.}~\bibnamefont {Wei}}, \bibinfo {author}
  {\bibfnamefont {S.}~\bibnamefont {Yang}}, \bibinfo {author} {\bibfnamefont
  {X.}~\bibnamefont {Zhang}}, \bibinfo {author} {\bibfnamefont {Y.-H.}\
  \bibnamefont {Zhang}}, \bibinfo {author} {\bibfnamefont {L.}~\bibnamefont
  {Zhao}},\ and\ \bibinfo {author} {\bibfnamefont {A.}~\bibnamefont {Zong}},\
  }\bibfield  {title} {\bibinfo {title} {\textit{The future of the correlated
  electron problem}},\ }\href {https://doi.org/10.21468/SciPostPhysCommRep.8}
  {\bibfield  {journal} {\bibinfo  {journal} {SciPost Phys. Comm. Rep.}\ ,\
  \bibinfo {pages} {8}} (\bibinfo {year} {2025})}\BibitemShut {NoStop}%
\bibitem [{\citenamefont {Arovas}\ \emph {et~al.}(2022)\citenamefont {Arovas},
  \citenamefont {Berg}, \citenamefont {Kivelson},\ and\ \citenamefont
  {Raghu}}]{arovas2022hubbard}%
  \BibitemOpen
  \bibfield  {author} {\bibinfo {author} {\bibfnamefont {D.~P.}\ \bibnamefont
  {Arovas}}, \bibinfo {author} {\bibfnamefont {E.}~\bibnamefont {Berg}},
  \bibinfo {author} {\bibfnamefont {S.~A.}\ \bibnamefont {Kivelson}},\ and\
  \bibinfo {author} {\bibfnamefont {S.}~\bibnamefont {Raghu}},\ }\bibfield
  {title} {\bibinfo {title} {\textit{The hubbard model}},\ }\href@noop {}
  {\bibfield  {journal} {\bibinfo  {journal} {Annual review of condensed matter
  physics}\ }\textbf {\bibinfo {volume} {13}},\ \bibinfo {pages} {239}
  (\bibinfo {year} {2022})}\BibitemShut {NoStop}%
\bibitem [{\citenamefont {Xu}\ \emph {et~al.}(2020)\citenamefont {Xu},
  \citenamefont {Liu}, \citenamefont {Rhodes}, \citenamefont {Watanabe},
  \citenamefont {Taniguchi}, \citenamefont {Hone}, \citenamefont {Elser},
  \citenamefont {Mak},\ and\ \citenamefont {Shan}}]{Xu2020}%
  \BibitemOpen
  \bibfield  {author} {\bibinfo {author} {\bibfnamefont {Y.}~\bibnamefont
  {Xu}}, \bibinfo {author} {\bibfnamefont {S.}~\bibnamefont {Liu}}, \bibinfo
  {author} {\bibfnamefont {D.~A.}\ \bibnamefont {Rhodes}}, \bibinfo {author}
  {\bibfnamefont {K.}~\bibnamefont {Watanabe}}, \bibinfo {author}
  {\bibfnamefont {T.}~\bibnamefont {Taniguchi}}, \bibinfo {author}
  {\bibfnamefont {J.}~\bibnamefont {Hone}}, \bibinfo {author} {\bibfnamefont
  {V.}~\bibnamefont {Elser}}, \bibinfo {author} {\bibfnamefont {K.~F.}\
  \bibnamefont {Mak}},\ and\ \bibinfo {author} {\bibfnamefont {J.}~\bibnamefont
  {Shan}},\ }\bibfield  {title} {\bibinfo {title} {\textit{Correlated
  insulating states at fractional fillings of moir{\'e} superlattices}},\
  }\href {https://doi.org/10.1038/s41586-020-2868-6} {\bibfield  {journal}
  {\bibinfo  {journal} {Nature}\ }\textbf {\bibinfo {volume} {587}},\ \bibinfo
  {pages} {214} (\bibinfo {year} {2020})}\BibitemShut {NoStop}%
\bibitem [{\citenamefont {Li}\ \emph {et~al.}(2021)\citenamefont {Li},
  \citenamefont {Li}, \citenamefont {Regan}, \citenamefont {Wang},
  \citenamefont {Zhao}, \citenamefont {Kahn}, \citenamefont {Yumigeta},
  \citenamefont {Blei}, \citenamefont {Taniguchi}, \citenamefont {Watanabe},
  \citenamefont {Tongay}, \citenamefont {Zettl}, \citenamefont {Crommie},\ and\
  \citenamefont {Wang}}]{Li2021}%
  \BibitemOpen
  \bibfield  {author} {\bibinfo {author} {\bibfnamefont {H.}~\bibnamefont
  {Li}}, \bibinfo {author} {\bibfnamefont {S.}~\bibnamefont {Li}}, \bibinfo
  {author} {\bibfnamefont {E.~C.}\ \bibnamefont {Regan}}, \bibinfo {author}
  {\bibfnamefont {D.}~\bibnamefont {Wang}}, \bibinfo {author} {\bibfnamefont
  {W.}~\bibnamefont {Zhao}}, \bibinfo {author} {\bibfnamefont {S.}~\bibnamefont
  {Kahn}}, \bibinfo {author} {\bibfnamefont {K.}~\bibnamefont {Yumigeta}},
  \bibinfo {author} {\bibfnamefont {M.}~\bibnamefont {Blei}}, \bibinfo {author}
  {\bibfnamefont {T.}~\bibnamefont {Taniguchi}}, \bibinfo {author}
  {\bibfnamefont {K.}~\bibnamefont {Watanabe}}, \bibinfo {author}
  {\bibfnamefont {S.}~\bibnamefont {Tongay}}, \bibinfo {author} {\bibfnamefont
  {A.}~\bibnamefont {Zettl}}, \bibinfo {author} {\bibfnamefont {M.~F.}\
  \bibnamefont {Crommie}},\ and\ \bibinfo {author} {\bibfnamefont
  {F.}~\bibnamefont {Wang}},\ }\bibfield  {title} {\bibinfo {title}
  {\textit{Imaging two-dimensional generalized Wigner crystals}},\ }\href
  {https://doi.org/10.1038/s41586-021-03874-9} {\bibfield  {journal} {\bibinfo
  {journal} {Nature}\ }\textbf {\bibinfo {volume} {597}},\ \bibinfo {pages}
  {650} (\bibinfo {year} {2021})}\BibitemShut {NoStop}%
\bibitem [{\citenamefont {Kumar}\ \emph {et~al.}(2025)\citenamefont {Kumar},
  \citenamefont {Lewandowski},\ and\ \citenamefont
  {Changlani}}]{kumar_wigner_2025}%
  \BibitemOpen
  \bibfield  {author} {\bibinfo {author} {\bibfnamefont {A.}~\bibnamefont
  {Kumar}}, \bibinfo {author} {\bibfnamefont {C.}~\bibnamefont {Lewandowski}},\
  and\ \bibinfo {author} {\bibfnamefont {H.~J.}\ \bibnamefont {Changlani}},\
  }\bibfield  {title} {\bibinfo {title} {\textit{Origin and stability of
  generalized Wigner crystallinity in triangular moir{\'e} systems}},\ }\href
  {https://doi.org/10.1038/s41535-025-00792-1} {\bibfield  {journal} {\bibinfo
  {journal} {npj Quantum Materials}\ }\textbf {\bibinfo {volume} {10}},\
  \bibinfo {pages} {95} (\bibinfo {year} {2025})}\BibitemShut {NoStop}%
\bibitem [{\citenamefont {Kim}\ \emph {et~al.}(2024)\citenamefont {Kim},
  \citenamefont {Esterlis}, \citenamefont {Murthy},\ and\ \citenamefont
  {Kivelson}}]{kim_kivelson_prb_2024}%
  \BibitemOpen
  \bibfield  {author} {\bibinfo {author} {\bibfnamefont {K.-S.}\ \bibnamefont
  {Kim}}, \bibinfo {author} {\bibfnamefont {I.}~\bibnamefont {Esterlis}},
  \bibinfo {author} {\bibfnamefont {C.}~\bibnamefont {Murthy}},\ and\ \bibinfo
  {author} {\bibfnamefont {S.~A.}\ \bibnamefont {Kivelson}},\ }\bibfield
  {title} {\bibinfo {title} {\textit{Dynamical defects in a two-dimensional
  Wigner crystal: Self-doping and kinetic magnetism}},\ }\href
  {https://doi.org/10.1103/PhysRevB.109.235130} {\bibfield  {journal} {\bibinfo
   {journal} {Phys. Rev. B}\ }\textbf {\bibinfo {volume} {109}},\ \bibinfo
  {pages} {235130} (\bibinfo {year} {2024})}\BibitemShut {NoStop}%
\bibitem [{\citenamefont {Tang}\ \emph {et~al.}(2020)\citenamefont {Tang},
  \citenamefont {Li}, \citenamefont {Li}, \citenamefont {Xu}, \citenamefont
  {Liu}, \citenamefont {Barmak}, \citenamefont {Watanabe}, \citenamefont
  {Taniguchi}, \citenamefont {MacDonald}, \citenamefont {Shan} \emph
  {et~al.}}]{tang2020simulation}%
  \BibitemOpen
  \bibfield  {author} {\bibinfo {author} {\bibfnamefont {Y.}~\bibnamefont
  {Tang}}, \bibinfo {author} {\bibfnamefont {L.}~\bibnamefont {Li}}, \bibinfo
  {author} {\bibfnamefont {T.}~\bibnamefont {Li}}, \bibinfo {author}
  {\bibfnamefont {Y.}~\bibnamefont {Xu}}, \bibinfo {author} {\bibfnamefont
  {S.}~\bibnamefont {Liu}}, \bibinfo {author} {\bibfnamefont {K.}~\bibnamefont
  {Barmak}}, \bibinfo {author} {\bibfnamefont {K.}~\bibnamefont {Watanabe}},
  \bibinfo {author} {\bibfnamefont {T.}~\bibnamefont {Taniguchi}}, \bibinfo
  {author} {\bibfnamefont {A.~H.}\ \bibnamefont {MacDonald}}, \bibinfo {author}
  {\bibfnamefont {J.}~\bibnamefont {Shan}}, \emph {et~al.},\ }\bibfield
  {title} {\bibinfo {title} {\textit{Simulation of Hubbard model physics in
  WSe2/WS2 moir{\'e} superlattices}},\ }\href
  {https://doi.org/https://doi.org/10.1038/s41586-020-2085-3} {\bibfield
  {journal} {\bibinfo  {journal} {Nature}\ }\textbf {\bibinfo {volume} {579}},\
  \bibinfo {pages} {353} (\bibinfo {year} {2020})}\BibitemShut {NoStop}%
\bibitem [{\citenamefont {LeBlanc}\ \emph {et~al.}(2015)\citenamefont
  {LeBlanc}, \citenamefont {Antipov}, \citenamefont {Becca}, \citenamefont
  {Bulik}, \citenamefont {Chan}, \citenamefont {Chung}, \citenamefont {Deng},
  \citenamefont {Ferrero}, \citenamefont {Henderson}, \citenamefont
  {Jim\'enez-Hoyos}, \citenamefont {Kozik}, \citenamefont {Liu}, \citenamefont
  {Millis}, \citenamefont {Prokof'ev}, \citenamefont {Qin}, \citenamefont
  {Scuseria}, \citenamefont {Shi}, \citenamefont {Svistunov}, \citenamefont
  {Tocchio}, \citenamefont {Tupitsyn}, \citenamefont {White}, \citenamefont
  {Zhang}, \citenamefont {Zheng}, \citenamefont {Zhu},\ and\ \citenamefont
  {Gull}}]{Simons_Hubbard_2015}%
  \BibitemOpen
  \bibfield  {author} {\bibinfo {author} {\bibfnamefont {J.~P.~F.}\
  \bibnamefont {LeBlanc}}, \bibinfo {author} {\bibfnamefont {A.~E.}\
  \bibnamefont {Antipov}}, \bibinfo {author} {\bibfnamefont {F.}~\bibnamefont
  {Becca}}, \bibinfo {author} {\bibfnamefont {I.~W.}\ \bibnamefont {Bulik}},
  \bibinfo {author} {\bibfnamefont {G.~K.-L.}\ \bibnamefont {Chan}}, \bibinfo
  {author} {\bibfnamefont {C.-M.}\ \bibnamefont {Chung}}, \bibinfo {author}
  {\bibfnamefont {Y.}~\bibnamefont {Deng}}, \bibinfo {author} {\bibfnamefont
  {M.}~\bibnamefont {Ferrero}}, \bibinfo {author} {\bibfnamefont {T.~M.}\
  \bibnamefont {Henderson}}, \bibinfo {author} {\bibfnamefont {C.~A.}\
  \bibnamefont {Jim\'enez-Hoyos}}, \bibinfo {author} {\bibfnamefont
  {E.}~\bibnamefont {Kozik}}, \bibinfo {author} {\bibfnamefont {X.-W.}\
  \bibnamefont {Liu}}, \bibinfo {author} {\bibfnamefont {A.~J.}\ \bibnamefont
  {Millis}}, \bibinfo {author} {\bibfnamefont {N.~V.}\ \bibnamefont
  {Prokof'ev}}, \bibinfo {author} {\bibfnamefont {M.}~\bibnamefont {Qin}},
  \bibinfo {author} {\bibfnamefont {G.~E.}\ \bibnamefont {Scuseria}}, \bibinfo
  {author} {\bibfnamefont {H.}~\bibnamefont {Shi}}, \bibinfo {author}
  {\bibfnamefont {B.~V.}\ \bibnamefont {Svistunov}}, \bibinfo {author}
  {\bibfnamefont {L.~F.}\ \bibnamefont {Tocchio}}, \bibinfo {author}
  {\bibfnamefont {I.~S.}\ \bibnamefont {Tupitsyn}}, \bibinfo {author}
  {\bibfnamefont {S.~R.}\ \bibnamefont {White}}, \bibinfo {author}
  {\bibfnamefont {S.}~\bibnamefont {Zhang}}, \bibinfo {author} {\bibfnamefont
  {B.-X.}\ \bibnamefont {Zheng}}, \bibinfo {author} {\bibfnamefont
  {Z.}~\bibnamefont {Zhu}},\ and\ \bibinfo {author} {\bibfnamefont
  {E.}~\bibnamefont {Gull}} (\bibinfo {collaboration} {Simons Collaboration on
  the Many-Electron Problem}),\ }\bibfield  {title} {\bibinfo {title}
  {\textit{Solutions of the Two-Dimensional Hubbard Model: Benchmarks and
  Results from a Wide Range of Numerical Algorithms}},\ }\href
  {https://doi.org/10.1103/PhysRevX.5.041041} {\bibfield  {journal} {\bibinfo
  {journal} {Phys. Rev. X}\ }\textbf {\bibinfo {volume} {5}},\ \bibinfo {pages}
  {041041} (\bibinfo {year} {2015})}\BibitemShut {NoStop}%
\bibitem [{\citenamefont {Bernu}\ \emph {et~al.}(1994)\citenamefont {Bernu},
  \citenamefont {Lecheminant}, \citenamefont {Lhuillier},\ and\ \citenamefont
  {Pierre}}]{Bernu_1994}%
  \BibitemOpen
  \bibfield  {author} {\bibinfo {author} {\bibfnamefont {B.}~\bibnamefont
  {Bernu}}, \bibinfo {author} {\bibfnamefont {P.}~\bibnamefont {Lecheminant}},
  \bibinfo {author} {\bibfnamefont {C.}~\bibnamefont {Lhuillier}},\ and\
  \bibinfo {author} {\bibfnamefont {L.}~\bibnamefont {Pierre}},\ }\bibfield
  {title} {\bibinfo {title} {\textit{Exact spectra, spin susceptibilities, and
  order parameter of the quantum Heisenberg antiferromagnet on the triangular
  lattice}},\ }\href {https://doi.org/10.1103/PhysRevB.50.10048} {\bibfield
  {journal} {\bibinfo  {journal} {Phys. Rev. B}\ }\textbf {\bibinfo {volume}
  {50}},\ \bibinfo {pages} {10048} (\bibinfo {year} {1994})}\BibitemShut
  {NoStop}%
\bibitem [{\citenamefont {Fishman}\ \emph {et~al.}(2022)\citenamefont
  {Fishman}, \citenamefont {White},\ and\ \citenamefont
  {Stoudenmire}}]{ITensor}%
  \BibitemOpen
  \bibfield  {author} {\bibinfo {author} {\bibfnamefont {M.}~\bibnamefont
  {Fishman}}, \bibinfo {author} {\bibfnamefont {S.~R.}\ \bibnamefont {White}},\
  and\ \bibinfo {author} {\bibfnamefont {E.~M.}\ \bibnamefont {Stoudenmire}},\
  }\bibfield  {title} {\bibinfo {title} {\textit{The ITensor Software Library
  for Tensor Network Calculations}},\ }\href
  {https://doi.org/10.21468/SciPostPhysCodeb.4} {\bibfield  {journal} {\bibinfo
   {journal} {SciPost Phys. Codebases}\ ,\ \bibinfo {pages} {4}} (\bibinfo
  {year} {2022})}\BibitemShut {NoStop}%
\bibitem [{\citenamefont {Weinberg}\ and\ \citenamefont
  {Bukov}(2017)}]{quspin}%
  \BibitemOpen
  \bibfield  {author} {\bibinfo {author} {\bibfnamefont {P.}~\bibnamefont
  {Weinberg}}\ and\ \bibinfo {author} {\bibfnamefont {M.}~\bibnamefont
  {Bukov}},\ }\bibfield  {title} {\bibinfo {title} {\textit{QuSpin: a Python
  package for dynamics and exact diagonalisation of quantum many body systems
  part I: spin chains}},\ }\href {https://doi.org/10.21468/SciPostPhys.2.1.003}
  {\bibfield  {journal} {\bibinfo  {journal} {SciPost Phys.}\ }\textbf
  {\bibinfo {volume} {2}},\ \bibinfo {pages} {003} (\bibinfo {year}
  {2017})}\BibitemShut {NoStop}%
\end{thebibliography}
%
\noindent
\\
\noindent
\MakeUppercase{\textbf{Acknowledgments}}\\
We thank B. S. Shastry, V. Elser, V. Dobrosavljevic, C. Lewandowski, A. Bernevig, E. Huang, S. Hill, M. Shatruk and K. Fossez for insightful discussions and for pointing us to previous literature. \addt{We also thank Dr. J. Carlstr\"om and the anonymous referee for valuable suggestions during the review process.} 
\addtt{We thank the Research Computing Center (RCC) and the Planck cluster at Florida State University (FSU), and the Bridges-2 cluster at the Pittsburgh Supercomputing Center for computational resources.}
\\
\\
\noindent
\addtt{
\MakeUppercase{\textbf{Funding statement}}\\
This work was supported by the National Science Foundation (NSF) Grant No. DMR 2046570 and the National High Magnetic Field Laboratory (NHMFL).
NHMFL is supported by the NSF Cooperative Agreement No. DMR-2128556 and the State of Florida. S.S. acknowledges additional support from the Dean’s Award for Doctoral Excellence (DADE) scholarship at FSU. A.K. was supported by the Dirac fellowship at NHMFL. The computer time on Bridges-2 was obtained through allocation PHY240324 (Towards predictive modeling of strongly correlated quantum matter) from the Advanced Cyberinfrastructure Coordination Ecosystem: Services and Support (ACCESS) program, which is supported by U.S. National Science Foundation grants $\#$2138259, $\#$2138286, $\#$2138307, $\#$2137603, and $\#$2138296.
}

\addt{
\section*{Author contributions}
H.J.C. conceived the initial project, with critical inputs from all authors during the study. S.S. and P.S. carried out the DMRG calculations, and A.K. performed the ED calculations, under the supervision of H.J.C. S.S., P.S. and A.K. prepared the figures and S.S. and H.J.C. wrote the paper; all authors contributed to editing the manuscript.
All authors were involved in scientific discussions at all stages.
}

\addt{
\section*{Competing Interests}
The authors declare no competing interests.
}

\end{document}